\documentclass{aa}  

\usepackage{graphicx}
\usepackage{txfonts}
\usepackage{natbib}
\bibpunct{(}{)}{;}{a}{}{,} 

\begin{document} 

   \title{First detections of FS Canis Majoris stars in clusters}

   \subtitle{Evolutionary state as constrained by coeval massive stars\thanks{Based on observations collected at the European Organisation for Astronomical Research in the Southern Hemisphere, Chile, under program IDs 083.D-0765 \& 087.D-0957.}}

   \author{D. de la Fuente\inst{1}
          \and
          F. Najarro\inst{1}
          \and
          C. Trombley\inst{2}
          \and
          B. Davies\inst{3}
          \and
          D. F. Figer\inst{2}
          }

   \institute{Centro de Astrobiolog\'ia (CSIC/INTA), ctra. de Ajalvir km. 4, 28850 Torrej\'on de Ardoz, Madrid, Spain\\
              \email{delafuente@cab.inta-csic.es}
         \and
             Center for Detectors, Rochester Institute of Technology, 74 Lomb Memorial Drive, Rochester, NY 14623, USA
         \and
	     Astrophysics Research Institute, Liverpool John Moores University, 146 Brownlow Hill, Liverpool L3 5RF, UK
             }

   \date{Received 19 November 2014 / Accepted 23 december 2014}

 
  \abstract
   {FS CMa stars are low-luminosity objects showing the B[e] phenomenon whose evolutionary state remains a puzzle. These stars are surrounded by compact disks of warm dust of unknown origin. Hitherto, membership of FS CMa stars to coeval populations has never been confirmed.}
   {The discovery of low-luminosity line emitters in the young massive clusters Mercer 20 and Mercer 70 prompts us to investigate the nature of such objects. We intend to confirm membership to coeval populations in order to characterize these emission-line stars through the cluster properties.}
   {Based on ISAAC/VLT medium-resolution spectroscopy and NICMOS/HST photometry of massive cluster members, new characterizations of Mercer 20 and Mercer 70 are performed. Coevality of each cluster and membership of the newly-discovered B[e] objects are investigated using our observations as well as literature data of the surroundings. Infrared excess and narrow-band photometric properties of the B[e] stars are also studied.}
   {We confirm and classify 22 new cluster members, including Wolf-Rayet stars and blue hypergiants. Spectral types (O9-B1.5 V) and radial velocities of B[e] objects are compatible with the remaining cluster members, while emission features of \ion{Mg}{ii}, \ion{Fe}{ii]}, and \ion{[Fe}{ii]} are identified in their spectra. The ages of these stars are 4.5 and 6 Myr, and they show mild infrared excesses.}
   {We confirm the presence of FS CMa stars in the coeval populations of Mercer 20 and Mercer 70. We discuss the nature and evolutionary state of FS CMa stars, discarding a post-AGB nature and introducing a new hypothesis about mergers. A new search method for FS CMa candidates in young massive clusters based on narrow-band Paschen-$\alpha$ photometry is proposed and tested in photometric data of other clusters, yielding three new candidates.}


   \maketitle
%

\section{Introduction}  

In the 1970s, several authors \citep{geisel70,ciatti+74,allen-swings72,allen-swings76} noticed that a peculiar group of Be stars showed forbidden emission lines together with a strong infrared excess. Members of this group were later referred as ``B[e] stars'' \citep[see e.g.][]{klutz-swings77,zickgraf-schulteladbeck89,bergner+95}. After more than two decades of research, it became evident that the B[e] class actually formed a very heterogeneous group of objects in diverse evolutionary phases, ranging from pre-main sequence stars to early stages of planetary nebulae. Consequently, \citet{lamers+98} proposed to drop the term ``B[e] stars'' and coined the expression ``B[e] phenomenon'' to describe the common observational features of these objects.

\citet{lamers+98} established five subclasses of objects showing the B[e] phenomenon, namely: B[e] supergiants (sgB[e]), pre-main sequence B[e]-type stars (HAeB[e]), compact planetary nebulae (cPNB[e]), symbiotic B[e] binaries (SymB[e]), and unclassified B[e]-type stars (unclB[e]). The latter was a miscellaneous group of objects that these authors could not include in any of the four remaining subclasses. In some cases, objects were incorporated into the unclB[e] group simply when the lack of conclusive observational evidence made the classification unclear \citep[e.g. CD$-$42º11721;][]{borgesfernandes+06} or controversial \citep[a typical case is MWC 349A; see][and references therein]{strelnitski+13}. As expected, when new analyses of such objects were carried out, part of them were moved to another B[e] subclass \citep[see e.g.][]{borgesfernandes+03}.

However, other unclB[e] stars had well-determined observational parameters that undoubtedly prevented them from being allocated in a different B[e] subclass. Among these, the most extensively studied cases are FS CMa \citep[see e.g.][]{swings+80,sorrell89,israelian+96,dewinter-vandenancker97,muratorio+06} and HD 50138 \citep[e.g.][]{hutsemekers85,pogodin97,jaschek-andrillat98,borgesfernandes+09}. Briefly, each one of these objects is composed of a star undergoing photometric and spectroscopic irregular variations, a dense circumstellar dusty disk, and a high-velocity polar wind.
Since the HIPPARCOS satellite has allowed to calculate their distances accurately, FS CMa and HD 50138 have precise luminosity measurements \citep{vandenancker+98}, which placed them in the main-sequence band or slightly above. Neither of them showed phenomena that are related to other B[e] subtypes, specifically: Roche lobe overflow, high-excitation forbidden lines, or association to nebulae or star-forming regions. On the other hand, both stars, as well as other similar objects, presented a distinctive steep decrease in the mid-infrared flux \citep{sheikina+00,miroshnichenko+02,miroshnichenko+06}, implying that the dusty disk is compact and warm.

In light of the above explained evidence, \citet{miroshnichenko07} grouped these low-luminosity B[e] objects with warm dust together in a new class, taking FS CMa as the prototype. According to the definition established by this author, FS CMa stars are those that fulfill the following criteria:

\begin{itemize}
 \item Spectral types between O9 and A2.
 \item Presence of the B[e] phenomenon, i.e. forbidden emission lines along with large infrared excess.
 \item Sharp decrease in the mid-infrared excess for $\lambda \gtrsim 20 \mu m$.
 \item Luminosity range: $2.5 \lesssim log(L/L_\odot) \lesssim 4.5$.
 \item Location outside of star-forming regions.
\end{itemize}

Subsequent observations \citep{miroshnichenko+07,miroshnichenko+09,miroshnichenko+11a,miroshnichenko+11b,borgesfernandes+09,borgesfernandes+12,rodriguez+12,polster+12,liermann+14} and modeling \citep{carciofi+10,borgesfernandes+11} of FS CMa stars have not been capable of unraveling the enigmatic nature of these objects. One of the most puzzling issues refers to the observationally inferred mass-loss rates, which are 2-3 dex higher than predicted by wind theory for B-type main-sequence stars. \citet{miroshnichenko+00} used the optical \ion{H}{i} emission of the FS CMa star AS 78 to obtain a very crude estimate of $\dot M \approx 1.5 \times 10^{-6} M_\odot / yr$.
Later, \citet{carciofi+10} calculated a more accurate value of $\dot M \approx 2.7 \times 10^{-7} M_\odot / yr$ for the model of IRAS 00470+6429, taking into account the geometrical effects through a latitude-dependent mass-loss rate per solid angle, together with constraints in the latitudinal density profile that allowed dust formation in a disk. In contrast, the mass-loss theoretical recipe of \citet{vink+00,vink+01} yields $9 \times 10^{-9}$ and $2 \times 10^{-9} M_\odot / yr$ respectively for these two FS CMa stars, assuming solar metallicity. Moreover, if these objects were normal low-luminosity stars, actual mass loss should be even lower due to the so-called ``weak wind problem'' \citep{puls+96,puls+08,martins+05b}.

The difficulties for explaining such extreme mass loss and the presence of a dusty disk around single main-sequence stars have led to hypotheses involving binarity. As discussed by \citet{miroshnichenko07, miroshnichenko11} and \citet{miroshnichenko+06, miroshnichenko+13}, mainly two binary evolution scenarios may provoke the emergence of the B[e] phenomenon in low-luminosity stars. First, FS CMa stars might be close binaries that have recently undergone a short phase of mass ejection. Such phase may be due to a rapid occurence of Roche lobe overflow that stops when the donor mass is low enough \citep{wellstein+01,vanrensbergen+08}, or could be triggered by the periastron passage of the companion in a highly eccentric orbit \citep{millour+09}.
Second, FS CMa stars may be intermediate-mass post-AGB stars in the first stages of the planetary nebulae ejection, where the dynamical interaction with a main-sequence companion can trap a (at least termporarly) stable circumbinary disk of hot dust \citep{vanwinckel07}. Although only 30 \% of FS CMa stars show signs of binarity \citep{miroshnichenko07,miroshnichenko+11a}, this low binary detection rate could be due to relatively low masses and luminosities of the companions as well as small separations between components.

Also, none of the secured FS CMa stars has been confirmed as a member of a coeval population, which hampers the inquiry of their evolutionary state. Although the FS CMa star IRAS 00470+6429 is projected towards the Cas OB7 association \citep{miroshnichenko+07}, membership is unlikely due to the absence of a interstellar component of a \ion{Na}{i} line that is present in the majority of the Cas OB7 members \citep{miroshnichenko+09}.

In this context, discovery of FS CMa objects as part of coeval, codistant star groups is crucial for understanding the nature and evolutionary state of FS CMa stars, linking the observational properties of these objects with the characteristics of the host populations, especially regarding ages. In this paper, we report the first detections of FS CMa stars in two clusters, specifically Mercer 20 and Mercer 70 (hereafter Mc20 and Mc70). We also study these objects through the new approach of a joint analysis with the cluster populations.
Interestingly, the discovery we present below is carried out through observations in the Paschen-$\alpha$ line that were initially aimed at finding hot massive evolved cluster members. $P_\alpha$ observations have been already proven to be very effective to find such stars in clusters \citep{davies+12,mc81_1} or in the central region of the Milky Way \citep{mauerhan-cotera+10,dong+11}, however this method has never been used to locate lower luminosity emission-line objects like FS CMa stars.

    \begin{figure*}
      \centering
      \includegraphics[width=8.cm, bb=40 15 695 590, clip]{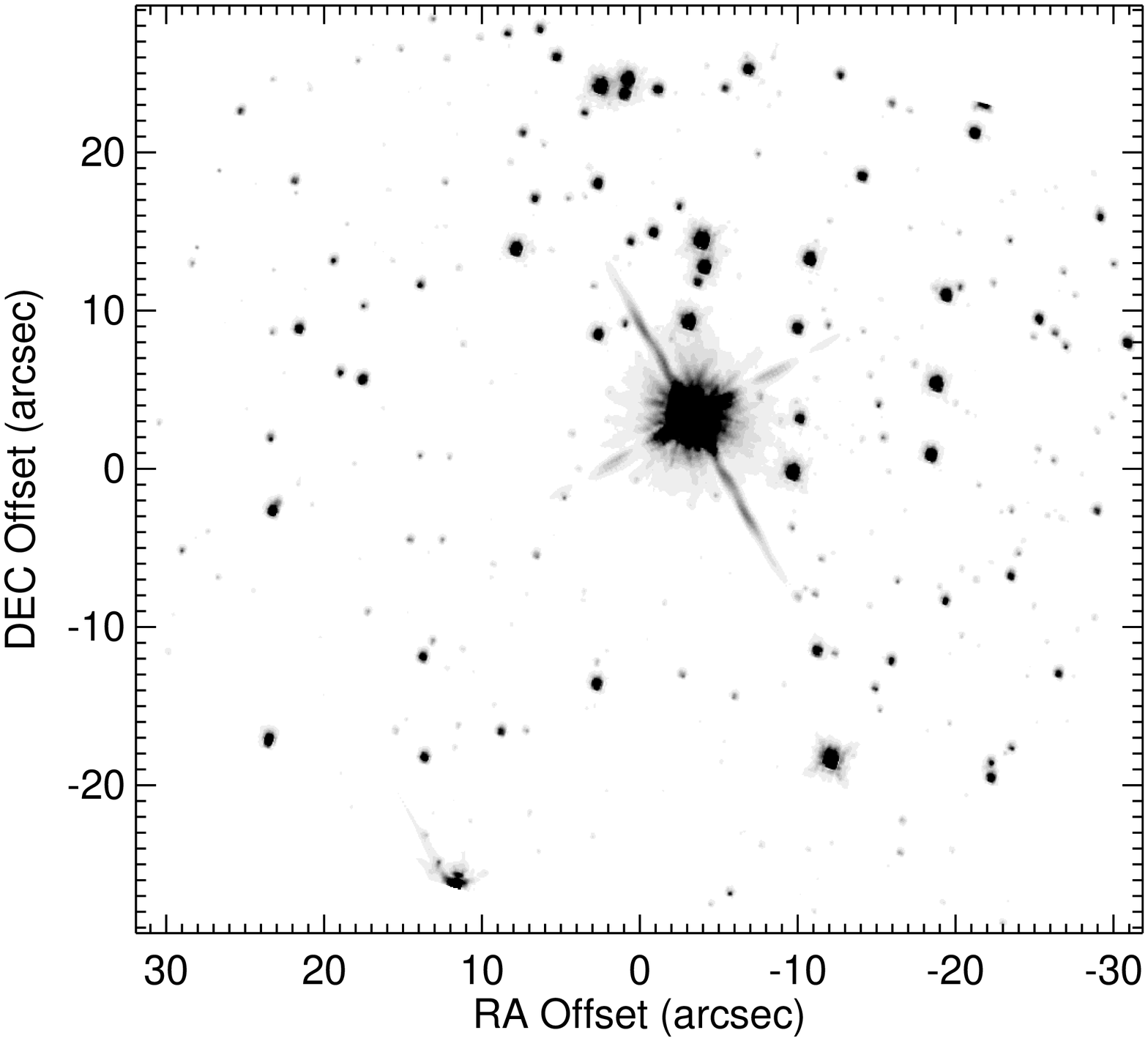}
      ~
      \includegraphics[width=8.cm, bb=20 15 675 590, clip]{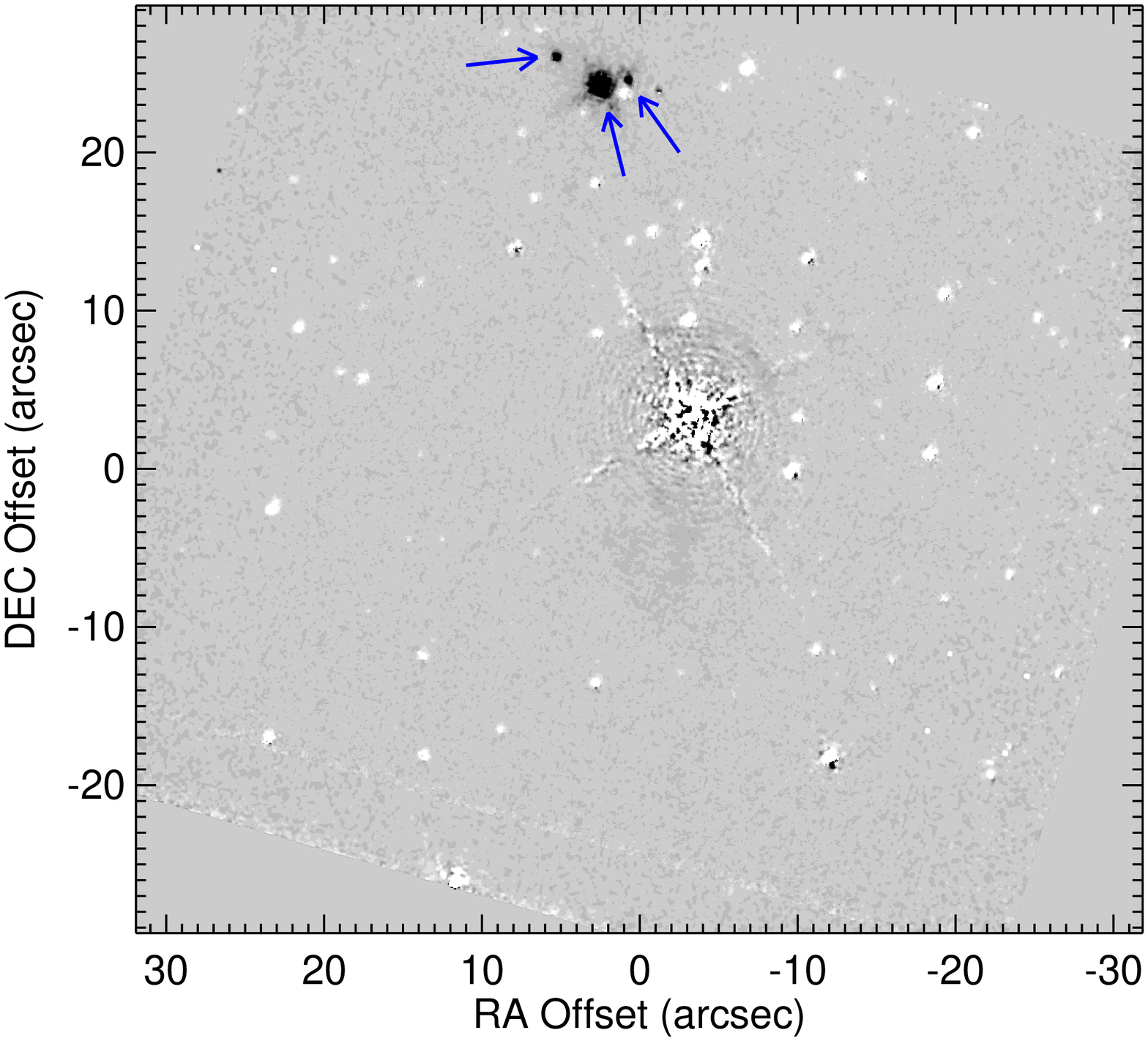}
      
      \caption{F222M image (\textit{left pannel}) and the F187N $-$ F190N subtraction (\textit{right pannel}) of Mercer 20. Strong $P_\alpha$ emitters are marked with blue arrows. Both images are centered at R.A. $ = 19^{h} 12^{m} 23.88^{s}$, Dec. $= +9\degr 57' 2.7''$}
      
      \label{fig:Mc20emissions}
    \end{figure*}

\section{Observations and data reduction}

  \subsection{Imaging}
  
 Clusters were observed on 2008 June 30 (Mc20) and 2008 July 12 (Mc70) with the NICMOS camera onboard the \textit{Hubble} Space Telescope (HST), as part of observing program \#11545 (PI: Davies). Near-infrared images were taken through filters F160W and F222M, as well as the narrow-band filters whose wavelengths correspond to $P_\alpha$ (P187N) and its adjacent continuum (P190N). Each frame has a field of view of $51.2'' \times 51.2''$ and a pixel scale of 0.203 arcsec/pix. Since resulting images and detailed photometry from this dataset were already published by \citet{trombley13}, we refer to that work for detailed description of data reduction.

    \begin{figure*}
      \centering
      \includegraphics[width=8.cm, bb=40 15 695 590, clip]{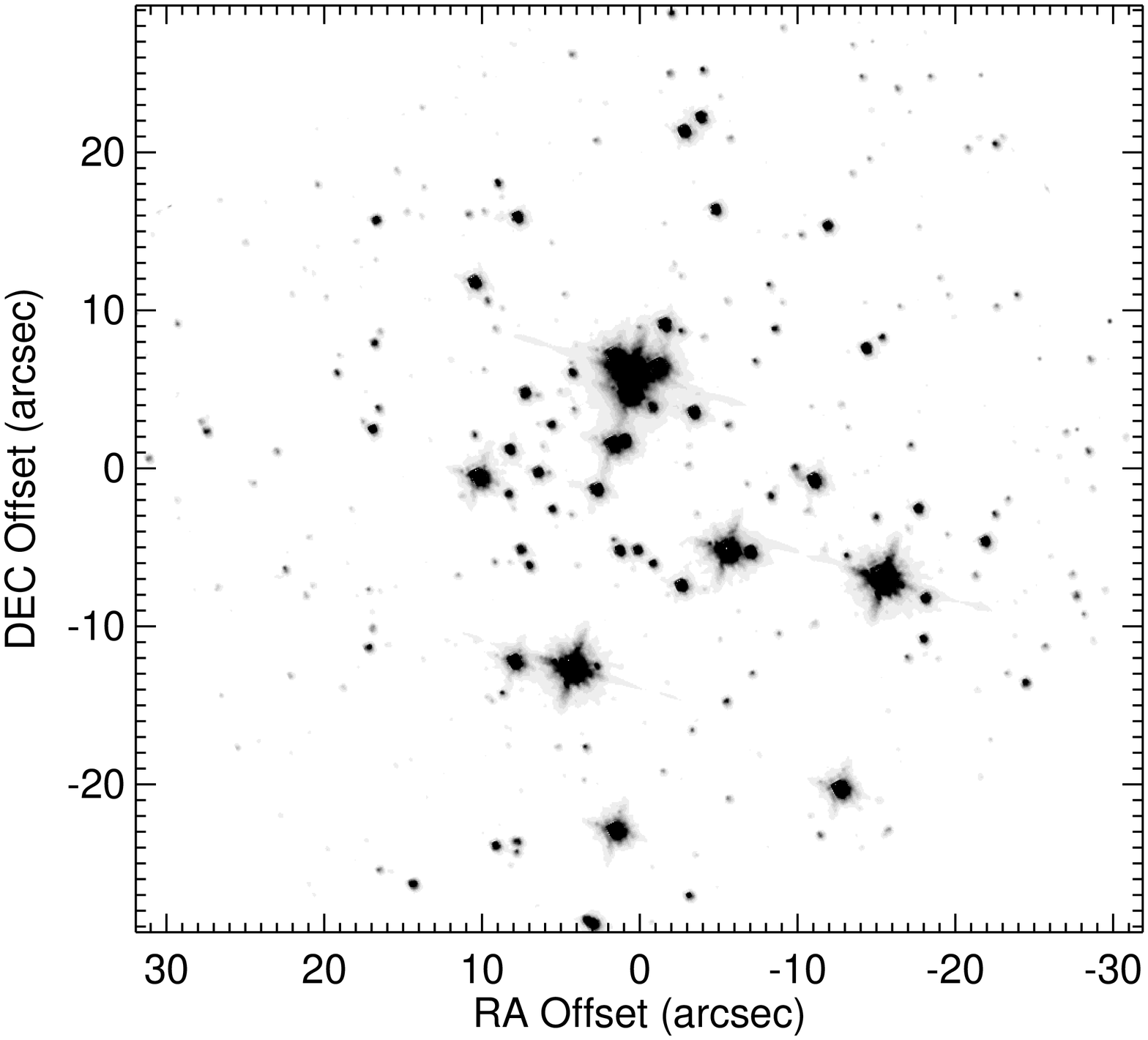}
      ~
      \includegraphics[width=8.cm, bb=20 15 675 590, clip]{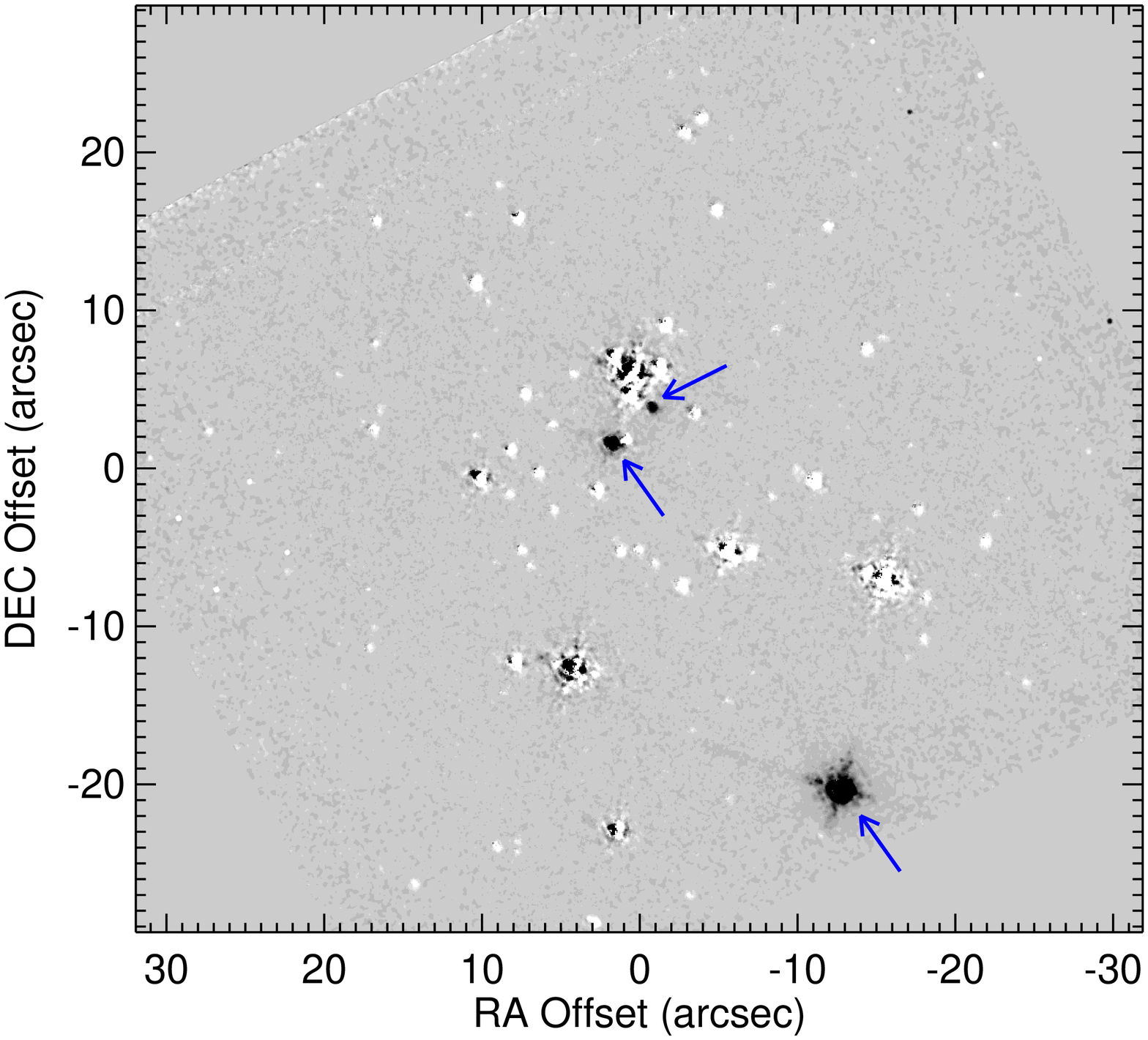}     
      \caption{Same as Fig. \ref{fig:Mc20emissions}, but for Mercer 70. Both images are centered at R.A. $ = 16^{h} 0^{m} 27.71^{s}$, Dec. $= -52\degr 10' 55.0''$}
      \label{fig:Mc70emissions}
    \end{figure*}

    \subsection{Target selection and spectroscopy}

For each cluster, we have built the difference image F187N $-$ F190N, which is intended to spotlight the objects with significant $P_\alpha$ emission, while the remaining sources are ``erased''. As seen in Figs. \ref{fig:Mc20emissions} and \ref{fig:Mc70emissions}, three strong $P_\alpha$ emitters are pinpointed in each cluster. We chose all six emission-line stars, as well as other bright targets in the central regions of Mc20 and Mc70, for spectroscopic follow-up.

Selected stars from Mc20 and Mc70 were observed with the ISAAC near-infrared spectrometer \citep{isaac} on the \textit{Very Large Telescope} (VLT), as part of service mode ESO programs 083.D-0765 (PI: Puga) and 087.D-0957 (PI: de la Fuente). Medium resolution, short wavelength spectroscopy was carried out with the the $0.8''$ slit, allowing a resolving power of $\lambda / {\Delta \lambda} \sim 4000$. Three different spectral settings were used: one in the H band, at central wavelength $\lambda_{cen} = 1.71 \mu m$; and two in the K band, at $\lambda_{cen} = 2.09 \mu m$ and $\lambda_{cen} = 2.21 \mu m$, hereafter K1 and K2 respectively. The observing time was optimized by orienting the slit in such a way that two targets of similar brightness were observed simultaneously.
Due to the crowded nature of the fields, additional stars were unintentionally situated within the slit, providing several bonus spectra. An ABBA nodding pattern was used for each slit position in order to remove the sky background. The nod shifts were carefully chosen to ensure that no stars are superimposed on the detector. The total integration times were computed so that a signal-to-noise ratio around 150 is reached for every programmed star. Additionally, observations of late-B, main sequence standard stars were requested in order to measure the telluric spectrum.
      
Reduction was carried out through a handmade, custom-built IDL pipeline that is described below. The first step consists of correcting the frame warping. The distortion along the spatial axis is estimated using the STARTRACE frames that are part of the ISAAC calibration plan. The OH emission spectrum that is imprinted in the science frames is used for calculating the distortion along the spectral axis as well as for the wavelength calibration. We have taken the wavelengths (in vacuum) of the OH emission lines from \citet{rousselot+00}. After rectification and calibration, the wavelength residuals have a RMS $\sim 0.5$ \AA. Sky background is removed from the calibrated frames through subtraction of each AB or BA nod pair.
All the positive and negative spectra of the resulting frames are then extracted and combined to obtain a 1-D spectrum for each object, including the telluric standards. As Brackett-series hydrogen lines are the only H- and K-band intrinsic features of our late-B standars, pure telluric spectra are obtained simply by removing such features by means of Voight-profile fitting. Each object spectrum is then divided by the corresponding telluric spectrum in order to get rid of the atmospheric absorption features. Finally, spectra are normalized through continuum fitting using 3rd to 4th degree polynomials.

      \begin{figure}
	\centering
	\includegraphics[width=0.85\hsize, bb=20 5 270 355, clip]{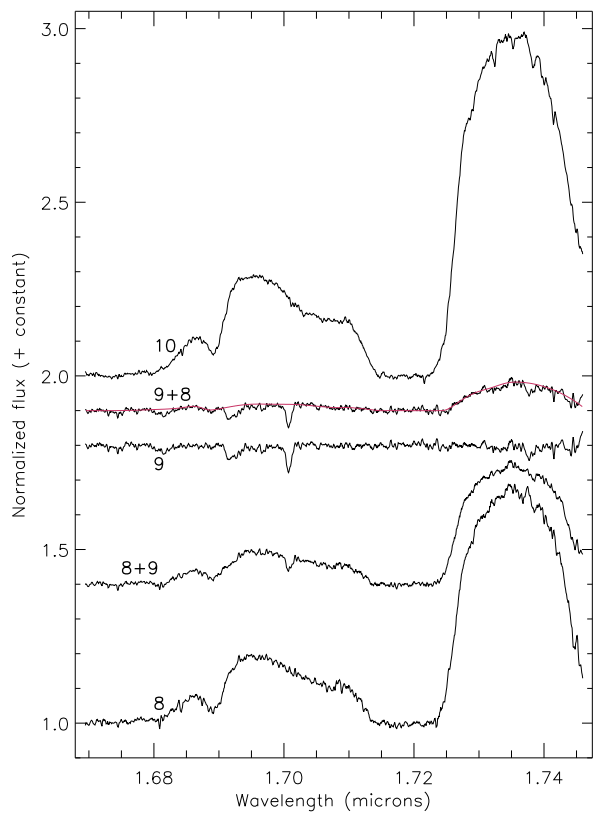}
	\caption{Plot illustrating the decontamination process for stars 8 and 9 of Mc20 in the H band. ``8+9'' and ``9+8'' are the input blended spectra, where the first addend is the star that contributes most. See text for details.
              }
        \label{fig:decontamination}
      \end{figure}

An additional step was necessary to separate the spectra Mc20-8 and Mc20-9, given that the small separation ($\approx 0.9 ''$) caused significant contamination between these two sources. The decontamination proccess, which is exemplified in Fig. \ref{fig:decontamination}, was possible due to the different morphologies of spectral features belonging to each star. While Mc20-9 is an O supergiant with narrow lines, Mc20-8 is a carbon-rich Wolf-Rayet star (WC) whose extremely broad emission lines barely change with spectral subtype \citep[see][]{figer+97}.
Taking the similar WC star Mc20-10 as a template, a cubic spline fitting (red line in Fig. \ref{fig:decontamination}) was made to subtract the Wolf-Rayet lines from the contaminated supergiant spectrum. The resulting Mc20-9 spectrum was then subtracted from the contaminated WC spectrum in order to obtain Mc20-8. Throughout the process, spectra were scaled properly according to the contribution of each individual spectrum to the blended ones, as measured through comparison of the line strengths.

\begin{table*}
  \caption{Equatorial coordinates, photometry, spectral types, radial velocities and membership of stars with new spectra in Mc20.}             
  \label{tab:mc20stars}      
  \centering          
  \begin{tabular}{c c c c c c c c c c c c}     
  \hline\hline
  Star & \multicolumn{2}{c}{Coordinates (J2000)} & \multicolumn{3}{c}{HST photometry} & \multicolumn{3}{c}{UKIDSS photometry} & Spectral& $\varv_r$ & Cluster\\
  ID 	& 	R.A. 	& 	Dec 	&	$m_{F160W}$ 	&	 $m_{F222M}$ 	& $P_\alpha$	& 	$J$	 & $H$	 & $K$ 		&  Type   & [km/s] & Member?\\ 
  \hline                    
   1 &  $19^h 12^m 23.61^s$   &   $9\degr 57' 37.4''$&\multicolumn{3}{c}{Out of the field of view}&11.37	& $-$ &	9.76	&B0-3 I &   $48$    & Yes \\
   2 &  $19^h 12^m 23.34^s$   &   $9\degr 57' 04.2''$	&11.44	&10.55 & -0.04	&12.52	&11.33&	10.45	&O9.5-B0 I-II& $61$ & Yes \\    
   3 &  $19^h 12^m 23.18^s$   &   $9\degr 56' 46.0''$	&10.53	&9.74  & -0.09	&11.47	&10.29&	9.49	&B2-4 I-II   & $19$ & Yes \\   
   4 &  $19^h 12^m 25.81^s$   &   $9\degr 57' 37.8''$&\multicolumn{3}{c}{Out of the field of view}&12.91	&11.62&	10.76	&O9 I &     $41$    & Yes \\ 
   5 &  $19^h 12^m 24.18^s$   &   $9\degr 57' 22.6''$	&13.23	&12.26 & -0.05	&14.29	&12.91&	12.06	&O9e       &  $58$  & Yes \\
   6 &  $19^h 12^m 23.72^s$   &   $9\degr 57' 19.0''$	&10.90	&10.05 & -0.05 	&\multicolumn{3}{c}{$-~$Unreliable$~-$}& O9-9.5 I & $24$ & Yes \\  
   7 &  $19^h 12^m 22.73^s$   &   $9\degr 57' 09.8''$	&11.79	&10.86 & -0.01	&12.77	&11.54&	10.73	&O9 II      & $34$  & Yes \\
   8 &  $19^h 12^m 24.05^s$   &   $9\degr 57' 29.1''$	&11.84	&10.66 & -0.19&\multicolumn{3}{c}{$-~$Unreliable$~-$}&WC 5-7  & ? & Yes \\
   9 &  $19^h 12^m 24.06^s$   &   $9\degr 57' 28.2''$	&12.52	&11.45 & -0.05&\multicolumn{3}{c}{$-~$Unreliable$~-$}&O6 If& $48$ & Yes \\
  10 &  $19^h 12^m 24.17^s$   &   $9\degr 57' 28.6''$	&11.22	&9.96  & -0.96&\multicolumn{3}{c}{$-~$Unreliable$~-$}&WC 5-7 &    ?   & Yes \\ 
  11 &  $19^h 12^m 22.68^s$   &   $9\degr 57' 15.5''$	&12.45	&11.55 & -0.03	&13.53	&12.12&	11.26	&O8.5-9 II &  $43$  & Yes \\    
  12 &  $19^h 12^m 22.74^s$   &   $9\degr 57' 05.3''$	&12.25	&11.44 & -0.03	&13.26	&12.02&	11.27	&O9 III   &   $34$  & Yes \\
  13 &  $19^h 12^m 23.53^s$   &   $9\degr 57' 29.7''$	&12.93	&11.14 & -0.03	&15.52	&12.53&	10.95	&KM    &    $63$    & No \\
  14 &  $19^h 12^m 24.93^s$   &   $9\degr 56' 52.6''$	&13.71	&12.19 & -0.02	&15.74	&13.33&	12.09	&KM    &    $-3$    & No \\
  15 &  $19^h 12^m 24.52^s$   &   $9\degr 57' 18.4''$	&11.98	&11.09 & -0.11	&13.25	&11.79&	10.89	&O5 If  &   $55$    & Yes \\
  16 &  $19^h 12^m 24.35^s$   &   $9\degr 57' 30.5''$	&13.55	&12.21 & -0.44	&14.84	&13.18&	11.91	&FS CMa  &   $45$   & Yes \\
  17 &  $19^h 12^m 23.29^s$   &   $9\degr 58' 44.7''$&\multicolumn{3}{c}{Out of the field of view}&14.59	&13.00&	12.08	&O9-B2 III-V & $51$ & Yes \\     
\hline                  
\end{tabular}
\end{table*}

\section{The host clusters}

The clusters were found by \citet{mercer+05} through an automated search of stellar overdensities in the GLIMPSE survey \citep{benjamin+03}. Further studies of Mc20 \citep{messineo+09, trombley13} and Mc70 \citep[also][]{trombley13} confirmed them as young clusters. Here we present a new analysis and characterization of both clusters.
    
  \begin{figure*}
    \centering
    \includegraphics[width=0.75\hsize, bb=0 5 394 585, clip]{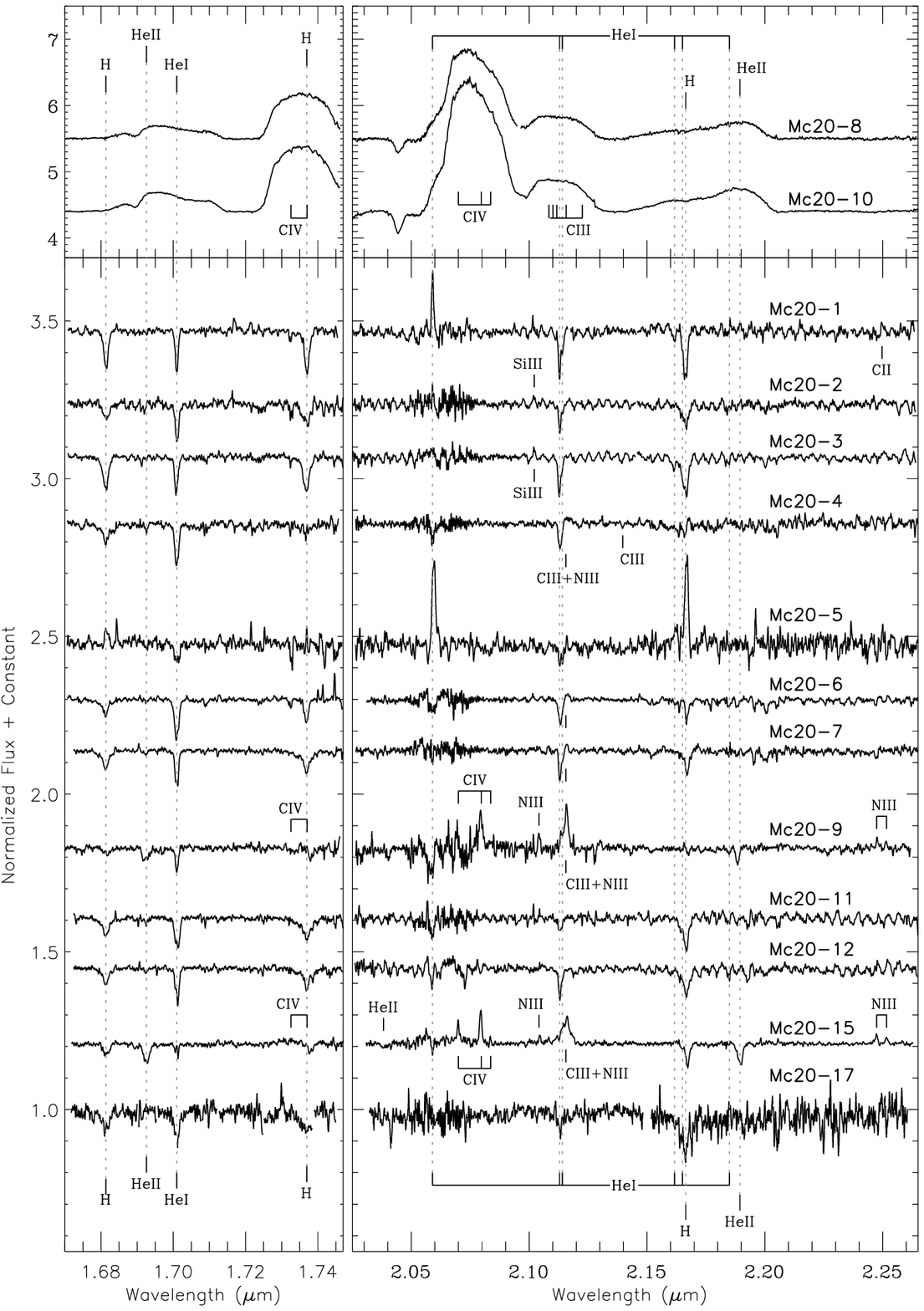}
    \caption{New H- and K-band spectra of Mc20 confirmed cluster members, except the FS CMa star. The upper chunk of the flux axis has been shrinked to show properly the strong lines of the WC stars. Spectral lines have been shifted to match the radial velocity of the cluster.
              }
    \label{fig:Mc20spectra}
  \end{figure*}

  \subsection{Stellar classification}
    \label{sec:typing}

\begin{figure*}
	\centering
    \includegraphics[width=0.75\hsize, bb=0 5 394 507, clip]{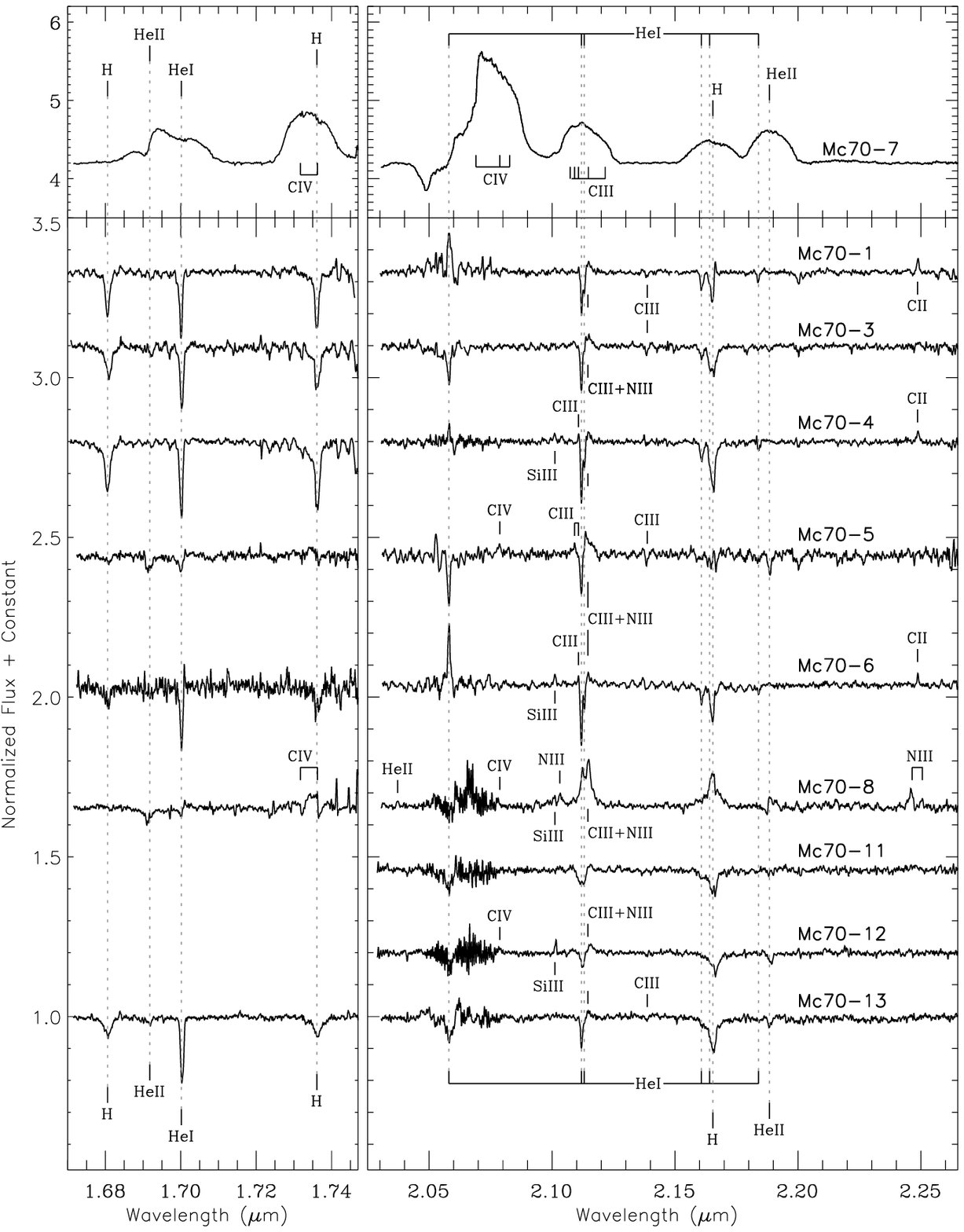}
    \caption{Same as Fig. \ref{fig:Mc20spectra}, but for Mercer 70.
              }
    \label{fig:Mc70spectra}
  \end{figure*}

\begin{figure*}
	\centering
    \includegraphics[width=0.75\hsize, bb=2 2 394 282, clip]{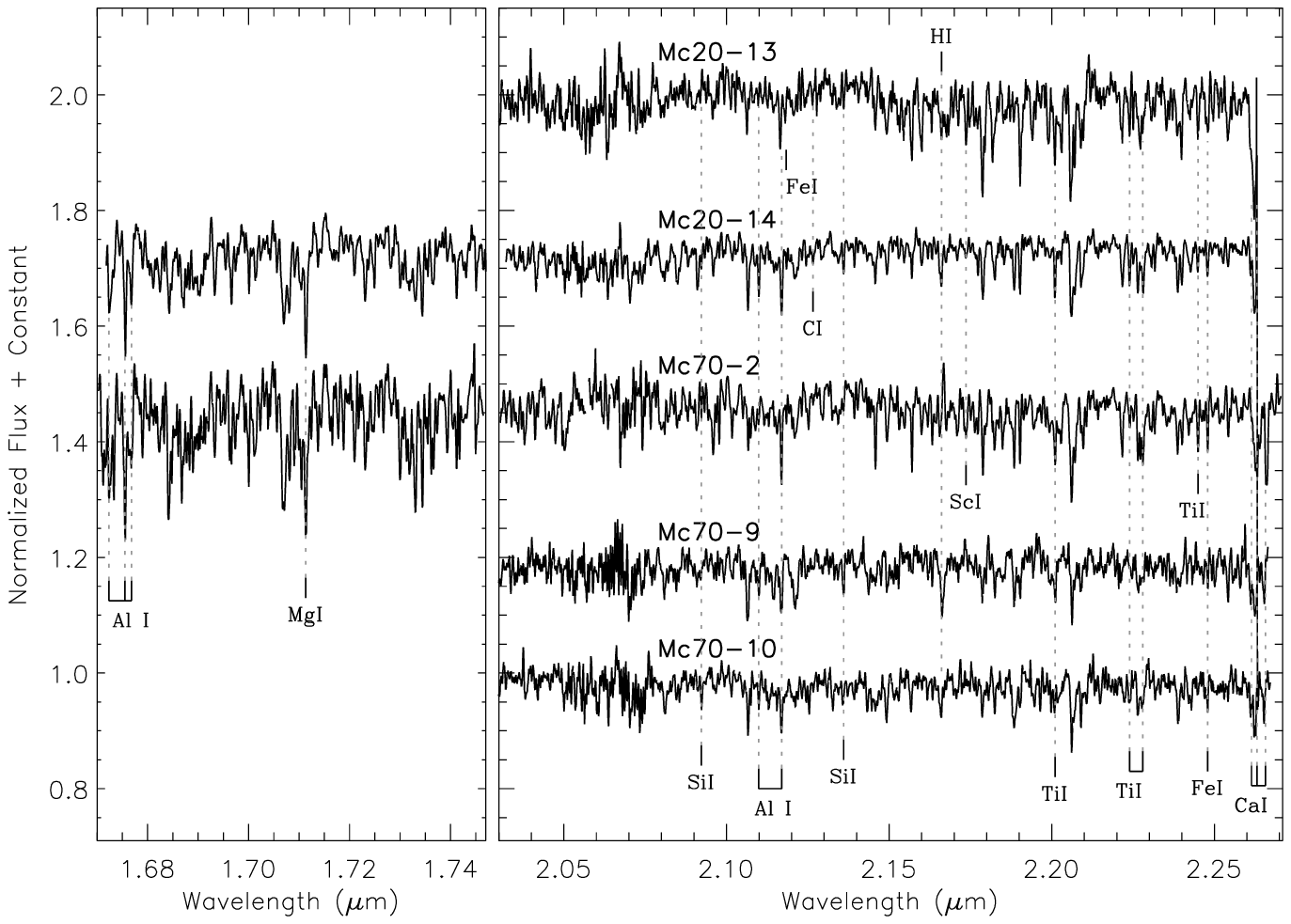}
    \caption{H- and K-band spectra of stars in the Mc20 and Mc70 fields that are not cluster members. Radial velocities have been corrected, therefore wavelengths are at rest.
              }
    \label{fig:coolspectra}
  \end{figure*}

\begin{figure}
	\centering
	\includegraphics[width=\hsize]{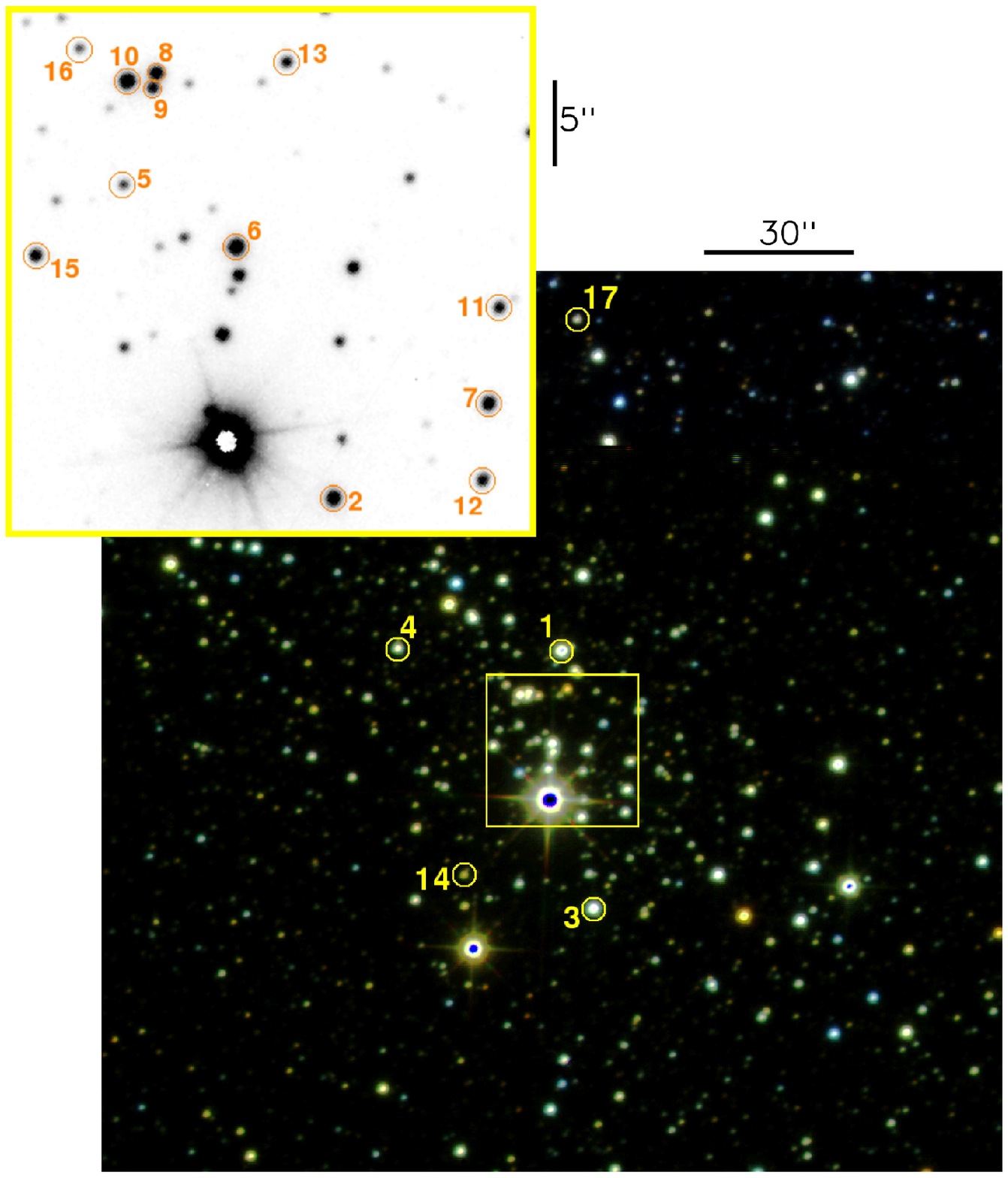}
	\caption{$3' \times 3'$ RGB image (R $=K$, G $=H$, B $=J$) of Mercer 20 (centered at R.A. $ = 19^{h} 12^{m} 23.70^{s}$, Dec. $= 9\degr 57' 23.6''$) from the UKIDSS survey, with a $30'' \times 30''$ close-up view of the central region (R.A. $ = 19^{h} 12^{m} 23.59^{s}$, Dec. $= 9\degr 57' 17.8''$) taken from a $1.71 \mu m$ narrow-band acquisition image of our ISAAC observations. North is up and east is left. All the stars with new spectra are identified, including foreground objects.
              }
        \label{fig:Mc20finders}
      \end{figure}

For every observed spectrum (Figs. \ref{fig:Mc20spectra}$-$\ref{fig:coolspectra}), line identification and stellar classification have been carried out using spectral atlases in H and K bands \citep{kleinmann-hall86, eenens+91, morris+96, hanson+96, hanson+98, hanson+05, figer+97, meyer+98, wallace-hinkle96, wallace-hinkle97, ivanov+04}, except for the FS CMa stars, which will be addressed in section \ref{sec:2fscma}. Final spectral types, together with all the information that is discussed in next subsections, are presented in Tables \ref{tab:mc20stars} and \ref{tab:mc70stars}. Stars are labeled as they appear in Figs. \ref{fig:Mc20finders} and \ref{fig:Mc70finders}.

\begin{table*}
  \caption{Equatorial coordinates, photometry, spectral types, radial velocities and membership of spectroscopically observed stars in Mc70}             
  \label{tab:mc70stars}      
  \centering          
  \begin{tabular}{c c c c c c c c c c c c}       
  \hline\hline       			
  Star & \multicolumn{2}{c}{Coordinates (J2000)} & \multicolumn{3}{c}{HST photometry} & Spectral& $\varv_r$ & Cluster\\
  ID 	& 	R.A. 	& 	Dec 	&	$m_{F160W}$ & $m_{F222M}$  & $P_\alpha$ &  Type   & [km/s] & Member?\\ 
  \hline
 1 &  $16^h 0^m 27.75^s$  &  $-52\degr 10' 44.39''$	&8.48 	&7.49  & -0.13	&B0-2 Ia+&   $-107$   	& Yes \\
 2 &  $16^h 0^m 26.00^s$  &  $-52\degr 10' 57.42''$	&8.67 	&7.69  & -0.06	&KM &     $-64$    	& No  \\ 
 3 &  $16^h 0^m 27.56^s$  &  $-52\degr 10' 44.03''$	&10.49	&9.63  & -0.07	&O9 I&     $-73$   	& Yes \\ 
 4 &  $16^h 0^m 27.08^s$  &  $-52\degr 10' 55.65''$	&9.77 	&8.59  & -0.06	&B0-2 I-II & $-88$ 	& Yes \\    
 5 &  $16^h 0^m 28.80^s$  &  $-52\degr 10' 51.02''$	&10.37	&9.64  & -0.09	&O6 I &     $-118$  	 & Yes \\
 6 &  $16^h 0^m 28.15^s$  &  $-52\degr 11'  3.13''$	&8.48 	&7.85  & -0.15	&B0-2 Ia+ &   $-92$  	& Yes \\ 
 7 &  $16^h 0^m 26.31^s$  &  $-52\degr 11' 10.73''$	&10.38	&9.45  & -0.93	&WC7   &    ?		 & Yes \\
 8 &  $16^h 0^m 27.86^s$  &  $-52\degr 10' 48.95''$	&10.64	&9.82  & -0.32	&OIfpe/WN   & ?		& Yes \\  
 9 &  $16^h 0^m 33.77^s$  &  $-52\degr 10' 54.68''$&\multicolumn{3}{c}{Out of the field of view}&G &$0$ & No  \\
10 &  $16^h 0^m 32.06^s$  &  $-52\degr 10' 53.31''$&\multicolumn{3}{c}{Out of the field of view}&KM&$-101$ & No  \\
11 &  $16^h 0^m 27.98^s$  &  $-52\degr 10' 51.60''$	&11.82	&11.03 & -0.05	&O9 III   &	$-101$& Yes \\     
12 &  $16^h 0^m 26.48^s$  &  $-52\degr 10' 51.20''$	&11.46	&10.81 & -0.06	&O6.5-8 III-V&	$-31$& Runaway \\      
13 &  $16^h 0^m 27.40^s$  &  $-52\degr 10' 57.90''$	&11.96	&11.37 & -0.04	&O7.5-8.5 III-IV&	$-75$& Yes \\      
14 &  $16^h 0^m 27.60^s$  &  $-52\degr 10' 46.66''$	&13.14	&12.11 & -0.44	&FS CMa      &	$-98$& Yes \\
\hline                  
\end{tabular}
\end{table*}

In particular, the temperature subtypes of OB stars have been distinguished through the \ion{He}{ii} / \ion{He}{i} strength ratio, especially in the H band, together with the existence of specific metallic lines in the K band. The luminosity classes are based on the width of the hydrogen lines, and the presence of emission components in H and He lines for the most luminous stars. On the other hand, the Wolf-Rayet stars are clearly differentiated by means of their extremely broad emission lines.

Cool stars have been harder to classify using our spectroscopic data. The only reliable way of finding the luminosity classes for these objects in the near infrared are the CO bandheads that are situated beyond $2.29 \mu m$, which are not covered by our spectroscopic span. Therefore, we only present a rough classification of cool stars based on the presence or absence of hydrogen lines and the ratio between the K-band \ion{Ca}{i} and \ion{Na}{i} lines.

Apart from spectral types listed on Table \ref{tab:mc20stars}, we have taken into account the SofI spectroscopic data of \citet{messineo+09}, which overlap our observations only partially. On the one hand, these authors studied the two following cluster members that were not covered by our observations. GLIMPSE20-1 is a Yellow Supergiant (YSG) of spectral type G0-2 I that dominates the infrared luminosity of the cluster. GLIMPSE20-9 is a OB star whose specific spectral subtype could not be determined due to a very low SNR.
On the other hand, the shared targets are GLIMPSE20-3 (= Mc20-1), GLIMPSE20-4 (= Mc20-6), GLIMPSE20-8 (= Mc20-3) and GLIMPSE20-6. Despite the latter was classified as an individual WC4-7 star by \citet{messineo+09}, our better spatial resolution has allowed us to resolve this object in a close group of two WC stars (Mc20-8 and Mc20-10) and an Of supergiant (Mc20-9). For all these overlapping sources, spectral types as determined in the present paper are preferred, given that the better quality of ISAAC spectra allows a more accurate classification.

Among our Mc70 spectroscopic targets, the only object with a cross-identification in the literature is Mc70-7 (= WR1038-22L). \citet{shara+12} found a WC7 classification based on a K-band spectrum of this object.

    \subsection{Radial velocities}
    
Radial velocities were calculated by means of gaussian fitting of symmetrical, unblended spectral lines whose wavelength accuracies are lower than one tenth of the spectral resolution element. Wavelengths in vacuum have been taken from the Van Hoof's Atomic Line List \footnote{http://www.pa.uky.edu/$\sim$peter/atomic}. Lines of which profiles show noticeable wind contamination have been discarded, given that their peaks appear shifted due to geometrical effects. When several measurements for the same star are available, results are averaged. In the worst cases (e.g. the Wolf-Rayet stars), there are no suitable lines for gaussian fitting, therefore radial velocity cannot be calculated using this method; ongoing atmosphere modelling (de la Fuente et al., in prep.) will allow us to solve this problem.

In order to estimate the error associated to this method, we have calculated the standard deviation of the residuals, considering only stars with at least 4 radial velocity measurements. This yields an uncertainty of 7 $km~s^{-1}$, which does not change if we take into account only early or late-type stars. Since this quantity is equivalent to one tenth of a resolution element, this demonstrates the robustness of our method and the accuracy of the wavelength calibration.

The resulting velocities have been corrected for the observer's contribution using the \textit{rvcorrect} routine from IRAF. We present both radial velocities (in Tables \ref{tab:mc20stars} and \ref{tab:mc70stars}) and spectra of cluster members (Figs. \ref{fig:Mc20spectra} and \ref{fig:Mc70spectra}) in the Local Standard of Rest (LSR) reference frame. Spectra of stars that are not cluster members (see section \ref{sec:membership}) are shown in \ref{fig:coolspectra}, where wavelengths are at rest.

The average values and standard deviations for each cluster are $\varv_{Mc20} = (43 \pm 4) ~km~s^{-1}$; $\sigma_{Mc20} = (13 \pm 3) ~km~s^{-1}$ and $\varv_{Mc70} = (-95 \pm 6) ~km~s^{-1}$; $\sigma_{Mc70} = (16 \pm 4) ~km~s^{-1}$. The Mc70 values have been calculated excluding the runaway star Mc70-12 (see discussion in section \ref{sec:membership}). After subtracting the measurement error ($7 ~km~s^{-1}$) in quadrature, we obtain $\sigma_{Mc20}^{disp} \approx 11 ~km~s^{-1}$ and $\sigma_{Mc70}^{disp} \approx 14 ~km~s^{-1}$ for the velocity dispersion of each cluster.

  \subsection{Photometry: a calibration between JHK and NICMOS magnitudes}
    \label{sec:photometry}

We have utilized the best available JHK photometric data of the observed stars. These data come from two different sources: the NICMOS/HST observations taken by \citet{trombley13}, and datasets from public surveys. UKIDSS \citep{ukidss} and VVV \citep{vvv} are normally the most suitable near-infrared public surveys for photometric studies of clusters in the northern and sourthern Galactic plane respectively, provided that the stellar density is not excessive. For Mc20, the UKIDSS catalog yields reliable JHK magnitudes for the spectroscopically observed objects, excepting four stars that are severely contaminated.
These UKIDSS magnitudes are consistent with their NICMOS counterparts, in such a way that $F160W$ and $F222M$ are always a fraction of magnitude greater than H and K. On the other hand, the Mc70 field is more problematic due to a significantly higher stellar density, which cause serious contamination in the majority of cluster members. Additionally, spectroscopically observed stars lie mostly in the nonlinear regime of VVV, which starts at $K \sim 11.5$ and $K \sim 12$ \citep{gonzalez+11, saito+12}. Consequently, we prefer to use photometric data from a public survey (UKIDSS) for the Mc20 case.

In principle, NICMOS provide the most suitable data due to a better spatial resolution, especially in highly crowded regions. However, absolute magnitudes and color indexes from the literature are usually given for Johnson-type (JHK) filter sets whose bandpasses differ significantly from the NICMOS photometric system. As we will show in section \ref{sec:extinction-distance}, these differences cause discrepant results, especially for the distance. As explained above, the JHK magnitudes are missing for Mc70, therefore a transformation between both photometric systems should be applied.
\citet{kim+05} presented such a transformation, whose validity range is $0.110 \le F160W-F222M \le 0.334$ (or alternatively, $0.046 \le J-K \le 0.242$). Mc70 stars have significantly larger color indexes due to extinction, therefore we have opted for calculating our own photometric transformation. We will take advantage of having photometric data available for Mc20 in both photometric systems to build a transformation between them.

      \begin{figure}
	\centering
	\includegraphics[width=\hsize]{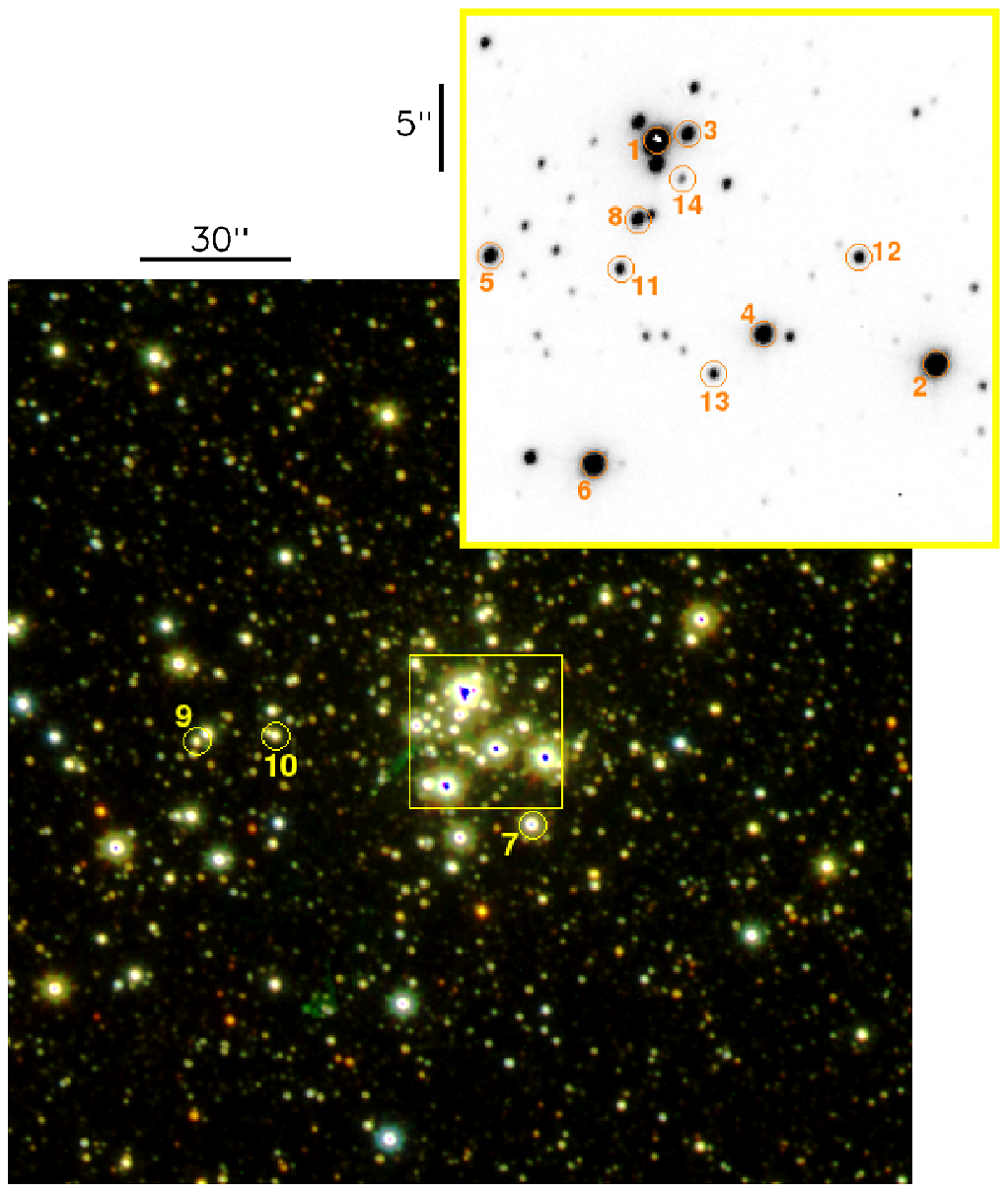}
	\caption{$3' \times 3'$ RGB image (R $=K$, G $=H$, B $=J$) of Mercer 70 (centered at R.A. $ = 16^{h} 0^{m} 28.00^{s}$, Dec. $= -52\degr 10' 51.4''$) from the VVV survey, with a $30'' \times 30''$ close-up view of the central region (R.A. $ = 16^{h} 0^{m} 27.42^{s}$, Dec. $= -52\degr 10' 51.7''$) taken from a $1.71 \mu m$ narrow-band acquisition image of our ISAAC observations. North is up and east is left. All the stars with new spectra are identified, including foreground objects.
	      }
	\label{fig:Mc70finders}
      \end{figure}

The equivalence between two photometric systems is a non-trivial problem, given that the transformations depend on the spectral types and reddening in a complicated manner. As discussed by \citet{stephens+00}, this is especially true when we compare JHK-type and NICMOS filters, due to significant differences in the wavelength ranges that are covered. However, if we only consider ``normal'' OB stars of Mc20 and Mc70 (i.e. ruling out stars with strong emission lines), spectral types are roughly similar (from O5 to B3), and color indexes are in a relatively narrow range ($ 0.6 \lesssim F160W - F222M \lesssim 1.0$, except a few stars that are excluded from this calculation). Therefore we can establish a simple transformation that is only valid for stars fulfilling such conditions. Using the Mc20 cluster members that have both NICMOS and UKIDSS photometry, the results are: $(F160W-H) = 0.22 \pm 0.07$ and $(F222M-K) = 0.19 \pm 0.07$.

  \subsection{Cluster membership}
  \label{sec:membership}

In order to discern which stars are cluster members, we have mainly relied on radial velocities and spectral types. As an additional criterion, color-based reddening is used with some flexibility, given that differential extinction can be present. When all these criteria are compatible with the majority of stars in the same field, they are considered cluster members; if none of these conditions are fulfilled, the object is surely a foreground or background star. All the spectroscopically observed stars are clearly settled in one of these categories, except two cases that we discuss below.

Mc70-10 is a late-type (K or M) star whose photometric data are unknown. Nevertheless, the VVV image of Mc70 (Fig. \ref{fig:Mc70finders}) shows this object as a relatively faint star, allowing to dismiss a red supergiant phase; otherwise it would be one of the brightest sources in the field. On the other hand, a red main sequence cluster member would be too faint to be observed, and the remaining evolutionary phases cannot be coeval with Wolf-Rayet (WR) and OB stars. Therefore Mc70-10 must be a foreground star.

Mc70-12 is a young star whose radial velocity differs in 64 $km~s^{-1}$ from the average of all the other cluster members, which corresponds to a $4.6 \sigma^{disp}$ difference. However, as the spectral type and magnitudes are congruent with the cluster population, it is unlikely to find such star being an unrelated object in the line of sight of the Mc70 central region. Also, this object has roughly the same $F160W - F222M$ color index than other O-type cluster members, therefore the extinction is similar. Hence, we infer Mc70-12 is probably a runaway star that was born in the cluster. Since this object is still appearing on the central region of Mc20, the ejection event must have occured very recently, unless its trajectory is nearly coincident with the line of sight.

  \subsection{Extinction and distance}
  \label{sec:extinction-distance}
  
In order to estimate the interstellar extiction in the directions of Mc20 and Mc70 as well as the heliocentric distances, we have used the photometric data of normal OB cluster members, the spectral types of which have well determined absolute magnitudes and color indexes. Mc20-9 and Mc70-4 have been excluded due to the abnormally high $F160W-F222M$ index of these stars, probably indicating additional circumstellar extinction. We have also dismissed the early B hypergiants Mc70-1 and Mc70-6, given that this kind of objects presents a large range of absolute magnitudes \citep{clark+12}. The intrinsic magnitudes and colors of the selected stars have been taken from \citet{straizys-kuriliene81,ducati+01,martins+05a,martins-plez06}. As explained above, these magnitudes and colors are tabulated in a JHK-type photometric system and for that reason we apply the transformations of section \ref{sec:photometry} to the NICMOS photometry of Mc70.

We assume an extinction law of the form $A_\lambda = \lambda^{-\alpha}$, where $\lambda$ can be approximated by the effective wavelengths of the UKIDSS filters \citep{hewett+06}. Several recent studies \citep{nishiyama+09,fitzpatrick-massa09,stead-hoare09,schoedel+10,fritz+11,wang-jiang14} have demonstrated such power law with exponents $\alpha \approx 2$ is suitable for the Galactic extinction in the near infrared. Since we can calculate the extinction index $A_K$ based on two different color indexes ($J-K$ and $H-K$) for Mc20, we can establish $\alpha$ as the value that matches both extinction calculations for each star.
We obtain an average of $\bar \alpha = 1.94 \pm 0.21$, which is in good agreement with the studies cited above. After using such value for both clusters, the average K-band extinction is $\bar A_K = 1.15 \pm 0.14$ for Mc20 and $\bar A_K = 1.01 \pm 0.14$ for Mc70. As a differential extinction cannot be discarded, we use the $A_K$ value for each star individually to calculate its corresponding distance. By averaging the distances, we obtain $d_{Mc20} = (8.2 \pm 1.3)~kpc$ and $d_{Mc70} = (7.0 \pm 0.9)~kpc$.

If we had directly used the $F160W - F222M$ color index, without further transformations, we would have found slightly lower extinction values ($\bar A_K = 1.10 \pm 0.08$ for Mc20 and $\bar A_K = 0.94 \pm 0.12$) and significantly farther distances ($d_{Mc20} = (9.7 \pm 1.7)~kpc$ and $d_{Mc70} = (7.9 \pm 1.0)~kpc$). The distance discrepancies, which are approximately equal than the uncertainties, justify the advisability of applying the photometric transformations.

\begin{figure}
	\centering
	\includegraphics[width=0.75\hsize, bb=10 5 270 240, clip]{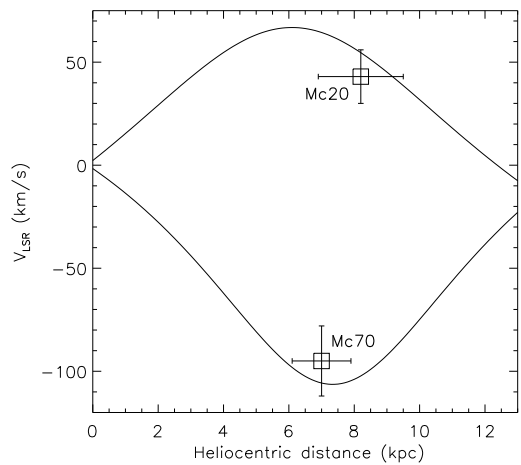}
	\caption{Distance-velocity diagram for Mercer 20 and Mercer 70, together with the projections of the Galactic rotation curve.
              }
        \label{fig:vgal}
      \end{figure}

In order to check the consistency of the distances with the radial velocities above calculated, we have superimposed both results on the projection of the Galactic rotation curve of \citet{brand-blitz93} along the Mc20/70 lines of sight, assuming a solar galactocentric distance of $8.5 kpc$. As shown in Fig. \ref{fig:vgal}, all these results are in good agreement. The possible radial velocity discrepancies ($\sim 10 ~km~s^{-1}$) are within the error bars, therefore we can neither confirm nor dismiss the existence of a peculiar velocity with respect to the Galaxy.

  \subsection{Age}
  \label{sec:age}
  
Since near-infrared color indexes are not good effective temperature indicators for hot stars, isochrone fitting of infrared color-magnitude diagrams is not suitable for age estimates of young clusters. We rely on the known spectral types instead, making them compatible with those expected from evolutionary models. Although \citet{messineo+09, trombley13} also made this work with Mc20 and Mc70, these authors used stellar grids that are now outdated.
The most recent Geneva evolutionary models at solar metallicity \citep{ekstrom+12,georgy+12} have implemented new opacities, reaction rates and mass-loss prescriptions, which cause significant changes in effective temperatures, luminosities and lifetimes of the different evolutionary stages, especially for rotating models. For what is of concern here, these new models allow WRs originate from less massive stars, therefore showing at older ages. Also, new rotating models change the expected ratios of WR subtypes and prevent Red Supergiants (RSGs) appear for $M_{ini} > 32 M_\odot$. New age determinations for Mc20 and Mc70 are discussed below, based on the above cited models and the associated isochrones and evolutionary tracks.

An upper limit for the ages can be established from a feature of the massive population that is shared by both clusters: the presence of WRs combined with the absence of RSGs. Given that Wolf-Rayet stars in Mc20 and Mc70 are really scarce (even if we include the Of/WN class), we cannot claim an age below the youngest RSG possible; we impose the weaker but more realistic condition RSG/WC$<1$. This occurs above $M_{ini} \sim 45 M_\odot$ for non-rotating models and $M_{ini} \sim 30 M_\odot$ for those with rotation, which correspond to age limits of 4.3 and 7.0 Myr respectively. In principle, the latter is preferred since rotating models are expected to be more realistic. However, initial masses near the lower aforementioned limits produce a WN/WC ratio well above 1, which is clearly against observational evidence in Mc20. As a consequence, the Mc20 age limit is slightly lowered to 6.5 Myr.

Another age constraint for Mc20 arises from the fact that the most luminous object is also the coolest one among the evolved stars. The less massive a star is, the lower the minimum effective temperature is reached during the post-main sequence lifetime. Therefore, the presence of a bright G0-2 supergiant ($log T_{eff} \approx 3.75$) constitutes an upper limit for the mass of the main sequence turn-off, which is equivalent to a lower limit for the cluster age. For non-rotating models, such temperatures are already reached for initial masses as high as $50 M_\odot$, which happens at an age of 4 Myr.
However, the resulting object would be too bright ($M_V=-9.9$) and the age must be increased. For the 5 Myr isochrone, magnitudes of the G0-2 supergiants are $M_V=-9.1$, which corresponds to the upper limit of the YSG luminosity in Mc20 within its error. Hence, a lower age limit of 5.0 Myr can be established from the non-rotating evolutionary tracks. Although using rotating models would require a considerably higher age ($\gtrsim 6.5$ Myr), we finally take the least restrictive limit since no information about rotation of GLIMPSE20-1 is available.

Regarding Mc70, the presence of two early-B hypergiants (eBHGs) provide strong constraints on the Mc70 main-sequence turn-off, as these objects are only expected for $40 M_\odot \lesssim M_{ini} \lesssim 60 M_\odot$ \citep{clark+12}. Taking into account both non-rotating and rotating tracks, this is equivalent to ages between 3.5 and 5.7 Moreover, from extinction and distance results above calculated, we obtain $M_V \approx -8.6$ for Mc70-1 and  $M_V \approx -8.3$ for Mc70-6; such magnitudes, together with their spectral type, turns these objects into analogs of the luminous hypergiant $\zeta ^1$ Sco \citep[B1.5 Ia+; $M_V=-8.93; log (L/L_{\odot})=6.10;$][]{clark+12}. Given that the $40 M_{\odot}$ Geneva tracks go through the corresponding effective temperatures significantly below $log (L/L_{\odot})= 6$, the turn-off cannot be close to such initial mass, therefore we take a lower limit of $45 M_\odot$, which is equivalent to an upper limit of 5.4 Myr.

In principle, the presence of mid-O supergiants in both clusters might provide additional upper limits for the age. However, the rotating $32 M_\odot$ isochrone is already reaching a turn-off luminosity of $log (L/L_{\odot})= 5.7$ with a mid-O temperature, and this occurs at 6.7 Myr. Such luminosity is compatible with the two mid-O supergiants in Mc20. Although this age constraint is less restrictive than the above found upper limits, the strong \ion{C}{iv} lines of Mc20-9 and Mc20-15 conflicts with the carbon abundance decline in that model.
Higher mases and lower ages would be required for rotating main-sequence stars ($\lesssim 5.5$ Myr) or for non-rotating models ($\lesssim 4.5$ Myr, which would be also in conflict with the existence of a YSG in Mc20). Since age determination of the mid-O stars is problematic (especially in Mc20) and dependent on metallicity, we prefer not to use these objects to constrain the age and to postpone this issue for future modeling that yield accurate stellar parameters, including abundances.

      \begin{figure}
	\centering
	\includegraphics[width=0.8\hsize, bb=30 0 330 320, clip]{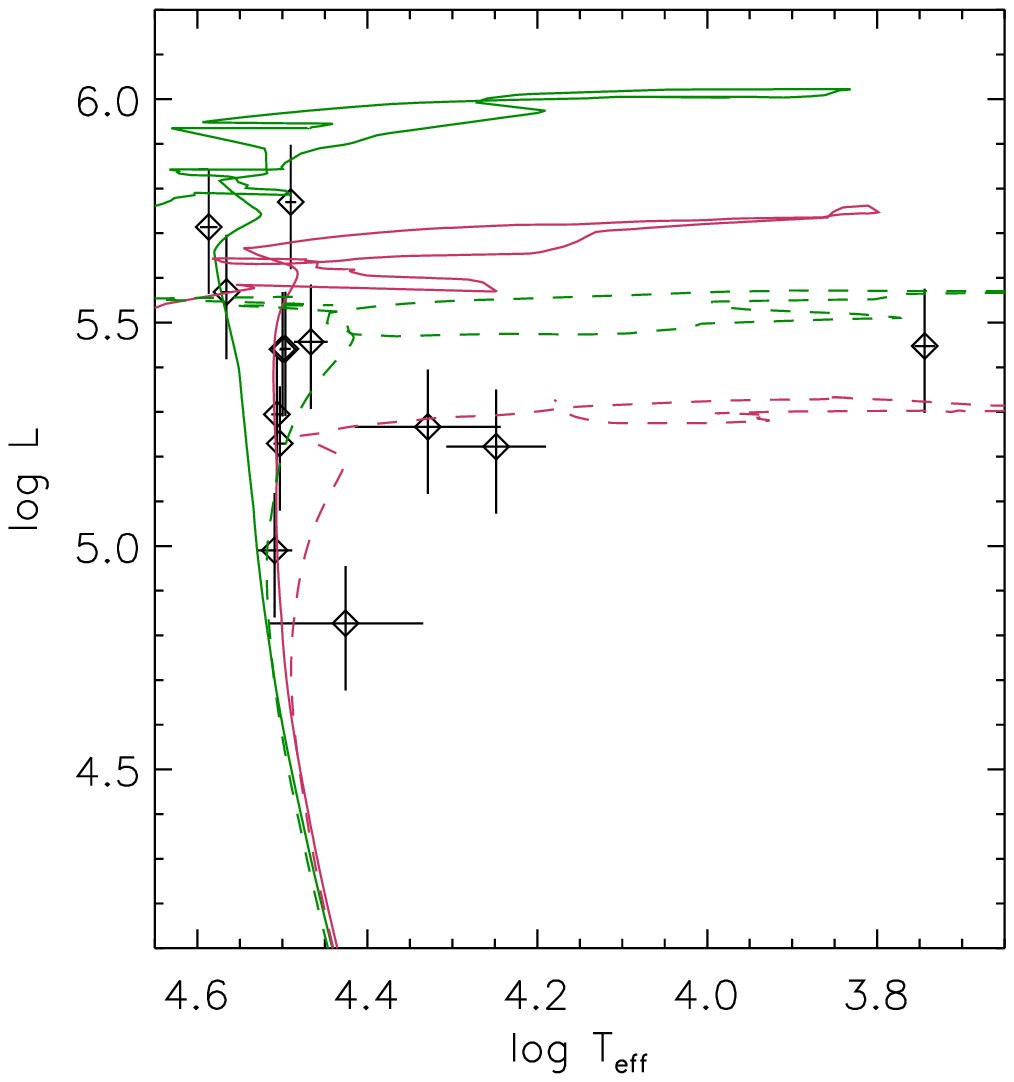}
	\includegraphics[width=0.8\hsize, bb=30 10 330 330, clip]{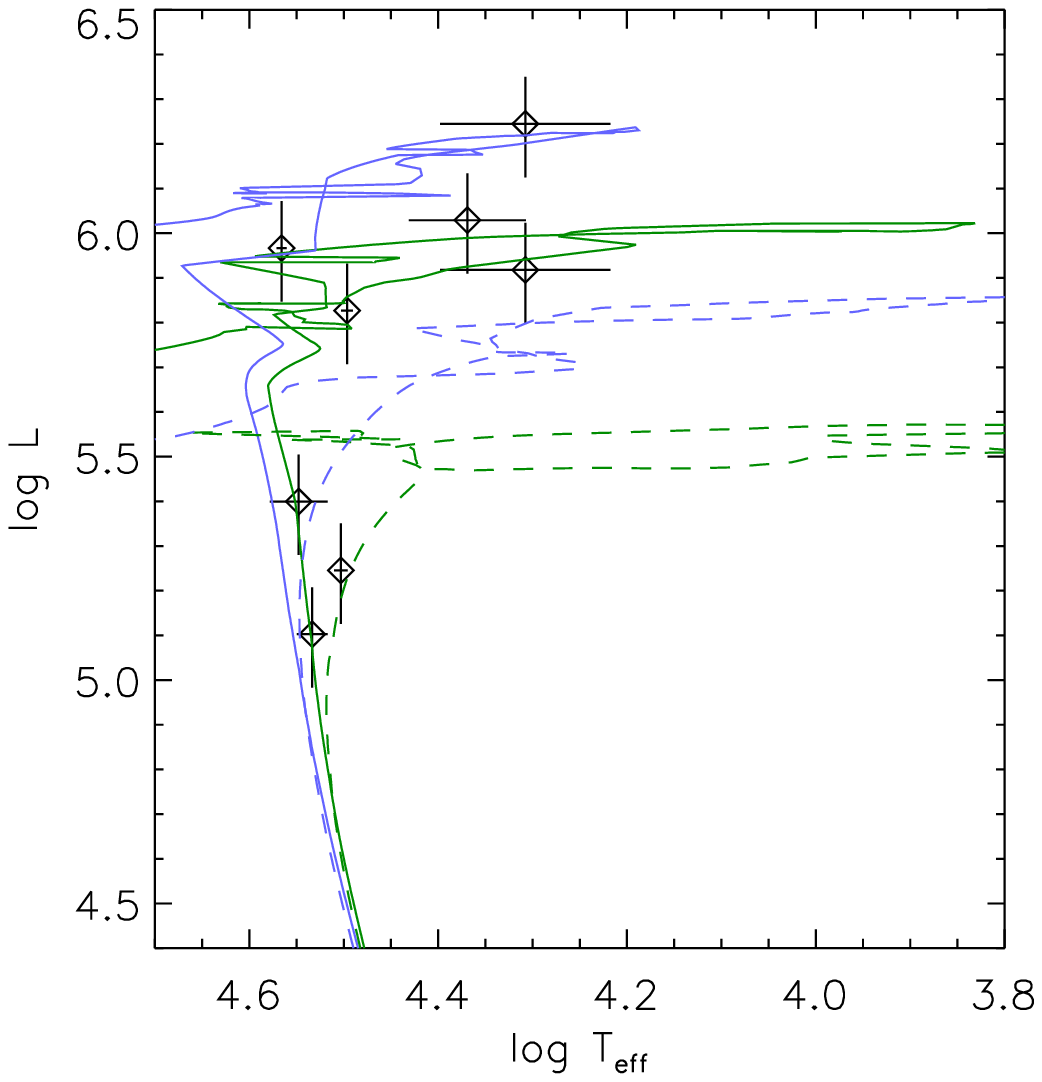}
	\caption{HR diagrams of Mercer 20 (\textit{top}) and Mercer 70 (\textit{bottom}), along with the rotating (solid line) and non-rotating (dashed) Geneva isochrones that are compatible with cluster ages, whose colors are as follows. Blue: 4.0 Myr; green: 5.0 Myr; red: 6.3 Myr. As the 5.0 Myr is coincidentally the lower limit of the Mc20 age, its population seems to fit closer to the 6.3 Myr isochrone.
              }
        \label{fig:hrd}
      \end{figure}

In summary, we estimate Mc20 is between 5.0 and 6.5 Myr old and the age of Mc70 is between 3.5 and 5.4 Myr. In order to illustrate our age derivation, we have built an HR diagram for each cluster showing the compatible Geneva isochrones (Fig. \ref{fig:hrd}). Effective temperatures are based on spectral types, using the temperature calibrations of \citet{martins+05a} (for O-type stars) and \citet{straizys-kuriliene81}. As these papers do not include hypergiant stars, we have based on \citet{clark+12} to find approximate values of the eBHGs temperatures, which are $\sim$ 15\% lower than for supergiants. We have used the tables of \citet{martins-plez06} (for O-type stars) and \citet{ducati+01}, along with the bolometric correction by \citet{torres10}, to find the luminosities. Since the circumstellar extinction and bolometric corrections for the emission-line stars (WR, B[e]) are highly uncertain, we have omitted these objects in Fig. \ref{fig:hrd}.

\subsection{Mass}

As the post-main sequence stellar populations of Mc20 and Mc70 are well sampled, we can use star counts in the upper region of the mass function to obtain a rough estimate of the total masses.

First of all, we need to establish the mass ranges that correspond to stars above the main sequence. Following the arguments in section \ref{sec:age}, the turn-off is situated at $M_{ini} \approx 32 M_\odot$ for Mc20 and $M_{ini} \approx 45 M_\odot$ for Mc70. The upper cut-off (i.e. the highest initial mass possible among stars that have not exploded as a supernova yet) is, however, very uncertain. Apart from variations associated with the age uncertainties, post-main sequence lifetimes of massive stars are largely dependent on rotation \citep{georgy+12}. Nonetheless, this limit can still be constrained to some extent, by means of the mass ratio of this upper end to the turn-off, which is around 1.5-2 for the suitable Geneva isochrones \citep{ekstrom+12}. After taking into account these considerations, we choose the following compromise values for the mass cut-off: $55 M_\odot$ for Mc20 and $80 M_\odot$ for Mc70.

The next step consists of counting post-main sequence stars. From the above cited evolutionary models, we derive turn-off luminosities of $log (L/L_\odot) \approx 5.6$ for Mc20 and   $log (L/L_\odot) \approx 5.8$ for Mc70. Every post-main sequence cluster member must be situated above these points, except a hypothetical low-luminosity WR star that still would be identified through its $P_\alpha$ emission. Using the O-type star calibrations by \citet{martins+05a,martins-plez06} for luminosity class III, along with extinction and distances towards Mc20 and Mc70, this is equivalent to the conditions $K_{post-MS} \lesssim 11$ and $K_{post-MS} \lesssim 10.3$ (or equivalently $F222M \lesssim 11.2$, $F222M \lesssim 10.5$), respectively. The latter is consistent with having the mid-O stars around the Mc20 turn-off (see section \ref{sec:age}).
After discarding stars with significantly redder colors ($H-K > 1$) than expected for hot luminous cluster members, we have found 9 additional stars (6 in Mc20 and 5 in Mc70) with no available spectra that fulfill these conditions. Among these, foreground stars could be present and must also be taken into account. Most of the foreground objects are easily identified as they usually have redder colors than non-OB luminous cluster members. From \citet{messineo+09} and the above presented data, seven foreground stars have been identified in the Mc20 field and three in the Mc70 region, however only one of them in each cluster (GLIMPSE20-11 and Mc70-2) have magnitudes and colors that could mimic evolved OB cluster members. In order to account for similar post-MS impostors in the sample of stars with no available spectra, we subtract $1 \pm 1$ stars in each cluster. After adding the spectroscopically confirmed post-MS members, we obtain $N_{Mc20} = 15 \pm 1$ and $N_{Mc20} = 11 \pm 1$.

In order to extrapolate these results to a wider mass range, we use the initial mass function of \citet{Salpeter55}. Integration between 0.5 and 150 $M_\odot$ give total masses of $M_{Mc20} = 1.3 \times 10^4$ and $M_{Mc70} = 1.5 \times 10^4$. Since the worst constrained parameter is the pre-supernova mass limit, we have estimated uncertainties through variations of this number within plausible values, yielding relative errors about 20\% for the cluster masses.

However, we should regard these results as upper limits as actual values could be noticeably lower. Apart from the difference between initial and current masses due to mass loss through stellar winds and supernovae, we have to consider that binary products (mass gainers and stellar mergers) are causing a significant overpopulation on the post-MS region where we have counted stars. As demonstrated by \citet{schneider+14}, such effect is expected to be substantial in massive clusters of a few Myr old. This is causing an overestimate of unknown extent in our mass values. Nevertheless, the Mc20 mass is expected to be well in excess of $3.4 \times 10^3 M_\odot$, which was the estimate of \citet{messineo+09}. These authors underestimated the number of stars above a specific luminosity limit owing to the lower derived distance and the existence of non-resolved massive stars (see section \ref{sec:typing}).

Alternatively, an upper limit for the mass can be established from the velocity dispersion, assuming that the clusters are thermalized and gravitationally bound. From the virial theorem, the total mass of such clusters can be approximated by $M_{cl} \approx 3 \sigma_{disp}^2 R / G$ \citep{ho-filippenko96,figer+02}, where $R$ is a representative radius of the sampled region. Taking $R_{Mc20} = 1 pc$ and $R_{Mc70} = 0.5 pc$, we obtain $M_{Mc20} \approx 8.4 \times 10^4 M_{\odot}$ and $M_{Mc70} \approx 6.8 \times 10^4 M_{\odot}$. These values are considerably higher than the above calculated upper limits.
At least part of this discrepancy could be caused by the high binary fraction that is expected for massive stars \citep{sana+12}. When velocity dispersion is measured through massive cluster members, binarity would cause dynamical masses estimates to exceed largely the photometrically calculated value, as demonstrated by \citet{gieles+10}. On the other hand, the assumptions of virial equilibrium and thermalization may not be true; Mc20 is particularly suspicious of being supervirial, since it is apparently very extended and quite loose in the UKIDSS images (Fig. \ref{fig:Mc20finders}).

\section{FS CMa stars in clusters}

In order to ensure that Mc20-16 and Mc70-14 are genuine FS CMa stars, we must verify these objects fulfill the definition of such stellar type \citep{miroshnichenko07}. This task is carried out in the present paper through two steps: first, proving the presence of the B[e] phenomenon, i.e. forbidden metallic emission lines together with IR excess; second, discarding other B[e] subtype classification. As we will discuss later, the second step can be performed considering that a low luminosity B[e] star that is several Myr old can only belong to the FS CMa subclass. Therefore, membership to the above studied clusters would be enough to secure such classification, provided that each cluster is coeval.

  \begin{figure*}
	\centering
    \includegraphics[width=\hsize]{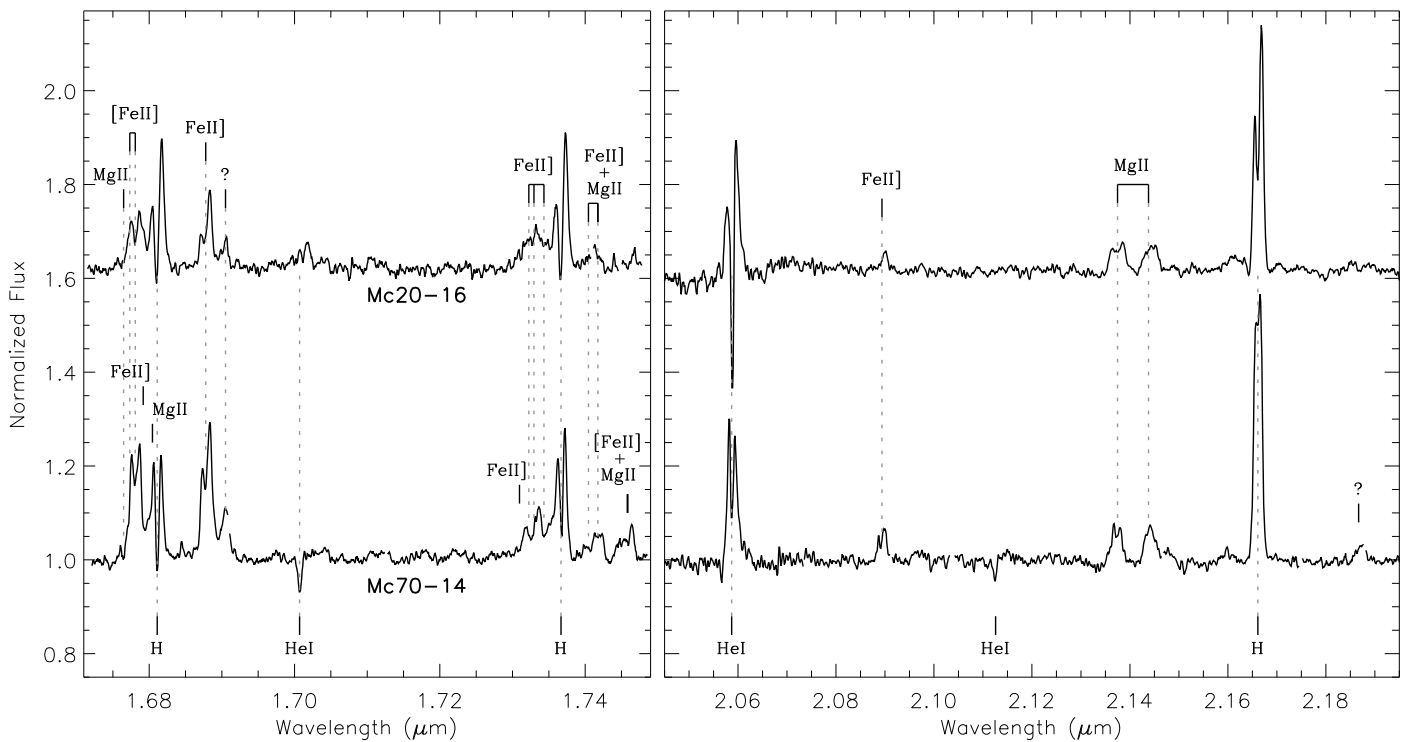}
    \caption{H- and K-band spectra of the FS CMa stars in Mc20 and Mc70. Unlike spectra in Figs. \ref{fig:Mc20spectra} and \ref{fig:Mc70spectra}, the wavelength axes are not relative to the LSR reference frame, but taking each star at rest. The wavelength region beyond $2.195 \mu m$ has been omitted since it only shows a featureless continuum for both stars.
              }
    \label{fig:fscmaspec}
  \end{figure*}

  \subsection{Spectra}
    \label{sec:2fscma}

Fig. \ref{fig:fscmaspec} shows spectra and line identification of the FS CMa stars we have detected in Mc20 and Mc70. Most of spectral features clearly present narrow, double-peaked emission line profiles which are typical of circumstellar disks. Such profiles are seen in the majority of the known FS CMa-type optical spectra \citep{miroshnichenko+07}, with exceptions probably having pole-on orientations. Strength differences between the two peaks, which are caused by disk oscillations \citep{okazaki91, hummel-hanuschik97}, are conspicuous in Mc20-16 and less evident in Mc70-14.
These asymmetries are normally quantified through the V(iolet)/R(ed) intensity ratio, as defined by \citet{Dachs+92}. For Mc20-16 lines,  $V/R \sim 1/2$, with extreme exceptions like the $2.089 \mu m$ line, showing a typical ``steeple shape'' \citep{hanuschik+95} where the weak peak almost disappears. $V/R$ is closer to unity in Mc70-14 lines; interestingly, the \ion{He}{i} $2.059 \mu m$ line presents a $V/R$ reversal phenomenon, which is likely due to the presence of a spiral density wave \citep{clark-steele00, wisniewski+07}.

\begin{table}
  \caption{Identification and violet/red radial velocities of detected spectral features on Mc20-16 and Mc70-14.}             
  \label{tab:detectedlines}      
  \centering          
  \begin{tabular}{c c c c c c}       
  \hline\hline       			
  $\lambda_{vacuum}$ & Ion 	& \multicolumn{2}{c}{$\varv_r^{Mc20-16}$ ($km/s$)}	& \multicolumn{2}{c}{$\varv_r^{Mc70-14}$ ($km/s$)}	\\
  (\AA) 	& 		&	viol./red 	&	mean 		&	 viol./red	&	mean   		\\
  \hline
  1.67648	& \ion{Mg}{ii}	&	-97/bld		&	?		&	-171/bld	&	?		\\
  1.67733	&\ion{[Fe}{ii]}	&	bld/bld		&	?		&	bld/bld		&	?		\\
  1.67809	&\ion{[Fe}{ii]}	& 	-56/146		&	45		& 	-187/6		&	-90.5		\\
  1.67918	& \ion{Fe}{ii]}	&	\multicolumn{2}{c}{non-detection}	&	bld/bld		& 	?		\\
  1.68045	& \ion{Mg}{ii}	&	\multicolumn{2}{c}{non-detection}	&	bld/bld		& 	?		\\
  1.68111	&   \ion{H}{i}	& 	-82/157		&	37.5		& 	-186/-5		&	-95.5		\\
  1.68778	& \ion{Fe}{ii]}	& 	-62/142		&	40		& 	-173/0		&	-86.5	\\
  $\sim$1.6905\tablefootmark{a}&?&	$-$		&	$-$		&	$-$		&	$-$		\\
  1.70070	&  \ion{He}{i}	&    \multicolumn{2}{c}{235\tablefootmark{b}}	&	(1 peak)	&	-95		\\
  1.73096	& \ion{Fe}{ii]}	&	\multicolumn{2}{c}{non-detection}	&	-170/bld	&	?		\\
  1.73225	& \ion{Fe}{ii]}	&	-79/bld		&	?		&	bld/bld		&	?		\\
  1.73295	& \ion{Fe}{ii]}	&	bld/bld		&	?		&	bld/bld		&	?		\\
  1.73432	& \ion{Fe}{ii]}	&	bld/135		&	?		&	bld/21		&	?		\\
  1.73668	&   \ion{H}{i}	& 	-79/155		&	38		& 	-174/-9		&	-91.5		\\
  1.74045	& \ion{Fe}{ii]}	&	-97/bld		&	?		&	-172/bld	&	?		\\
  1.74167	& \ion{Mg}{ii}	&	bld/bld		&	?		&	bld/bld		&	?		\\
  1.74188	& \ion{Fe}{ii]}	&	bld/bld		&	?		&	bld/bld		&	?		\\
  1.74541	&\ion{[Fe}{ii]}	&	\multicolumn{2}{c}{non-detection}	&	-214/tellu	&	?		\\
  1.74586	& \ion{Mg}{ii}	&	\multicolumn{2}{c}{non-detection}	&	tellu/-4	&	?		\\
  2.05869	&  \ion{He}{i}	& 	-85/179		&	47		& 	-173/0		&	-86.5		\\
  2.08938	& \ion{Fe}{ii]}	& 	-27/137		&	55		& 	-191/-29	&	-110		\\
  2.11258	&  \ion{He}{i}	&	\multicolumn{2}{c}{non-detection}	&	(1 peak)	&	-114		\\
  2.13748	& \ion{Mg}{ii}	& 	-99/190		&	45.5		& 	-200/-27	&	-113.5		\\
  2.14380	& \ion{Mg}{ii}	&	\multicolumn{2}{c}{unresolved peaks}	& 	-169/-47	&	-108		\\
  2.16612	&   \ion{H}{i}	& 	-39/143		&	52		& 	-141/-36	&	-88.5		\\
  $\sim$2.1867\tablefootmark{a}&?&	\multicolumn{2}{c}{non-detection}	&	$-$		&	$-$		\\
  \hline                  
\end{tabular}
\tablefoot{Unknown data are labeled as question marks. When radial velocities cannot be reliably calculated, short explanations are given, where the following abbreviations are used: bld, blended with a neighboring line; tellu, a strong telluric line distorts the peak shape.
\tablefoottext{a}{Vacuum wavelengths of unidentified lines have been estimated so that the radial velocities are roughly coincident with the values for the rest of lines.}
\tablefoottext{b}{This value is not considered for average radial velocity since the peak is neither photospheric nor disk-generated; see text for discussion.}}
\end{table}
    
Despite the predominant disk contribution, a few stellar spectral features, consisting of faint \ion{He}{i} lines, are also seen. The $1.701 \mu m$ and $2.113 \mu m$ lines appear only in narrow absorption in Mc70-14, therefore these are photospheric. The $1.701 \mu m $ absorption feature allows us to find an upper limit of $\varv \sin i \approx 70 ~km~s^{-1}$ for the projected rotational velocity, following the methodology of \citet{simondiaz-herrero14}. On the other hand, no absorptions are present in the Mc20-16 spectrum, but the $1.701 \mu m$ \ion{He}{i} line shows an asymmetric, remarkably red-shifted emission, which points at the existence of a high-velocity polar wind. The absorption component that this line should have is probably filled with disk emission of slightly higher intensity, as suggested by a bump on the blue wing of the wind emission.

Given that circumstellar matter has lower temperature than the photosperes in general, disk spectral features only give lower limits for the effective temperature of the underlying stars. In this respect, the \ion{He}{i} $2.059 \mu m$ line provides the strongest constraint, being expected in emission only for Be stars earlier than B2.5 \citep{clark-steele00}. However, further constraints arises from the aforementioned \ion{He}{i} stellar features, taking into account that such lines appear very weakened due to dominance of disk contribution.
The $1.701 \mu m$ and $2.113 \mu m$ absorptions decrease with B subtypes, being this change more abrupt for lower luminosities \citep{hanson+96, hanson+98}. In B2V stars, these lines would be faint enough to vanish in disk-dominated spectra, therefore Mc20-16 and Mc70-14 temperature types are B1.5 or earlier. Similarly, the higher temperature limit corresponds to an O9 subtype, as hypothetical \ion{He}{ii} features would be weak enough to be outshined by the disks. Finally, the presence of wind emission at $1.701 \mu m$ favours the earliest subtypes for Mc20-16 ($<$ B1).

As no H-band line identification work is available in the literature for FS CMa stars, we have taken spectroscopically similar spectra as a reference, namely early-B Luminous Blue Variables (LBVs) with forbidden lines. Specifically, we have relied on the medium resolution, high signal-to-noise near-infrared spectra of $\eta$ Carinae \citep{hamann+97}, the Pistol Star and qF 362 \citep{najarro+09}. Despite the great physical discrepancies (especially in terms of luminosity) between both stellar types \citep{conti97}, FS CMa-type spectra bear a striking resemblance to these LBVs. Single-peaked emission profiles are the only qualitative difference; luckily, this makes line identification easier, given that neighboring lines are separated better. Regarding the K band, we have also used the aforementioned LBV spectra in addition to the recent B[e] observations of \citet{liermann+14}.

Table \ref{tab:detectedlines} lists the wavelengths and radial velocities of every spectral feature that is detected in at least one of the FS CMa stars. Velocity calculations of the emission lines have been problematic due to their strong asymmetry. Central shell-type absorptions in double-lined profiles are not suitable for radial velocity measurements owing to strong contamination from the asymmetric disk emission, causing an apparent shift of the central peak toward the weaker emission component. The only reliable method we have found for estimating radial velocity of each double-lined peak consists of averaging the measurements of the violet and red peaks, which are also shown in Table \ref{tab:detectedlines}.
First, velocity determination has been carried out through gaussian fitting of each unblended or marginally blended peak, using only a few pixels around the maximum to avoid wing effects and contamination from the sibling peak. Second, we have used the mirroring method \citep{parimucha-skoda07} for the same peak regions in order to test the validity of our approach. The maximum discrepance between both methods is $4 ~km~s^{-1}$, which is well below one tenth of the resolution element. Final average velocities for the B[e] objects are $\varv_{Mc20-16} = (45 \pm 6) ~km~s^{-1}$ and $\varv_{Mc70-14} = (-98 \pm 11) ~km~s^{-1}$. These results are in good agreement with Mc20 and Mc70 radial velocities, respectively.

Since each double-peaked emission feature samples the disk region where the line is formed, the velocity field can be mapped through peak separations. For this purpose, comparison between Brackett-series hydrogen lines is particularly useful, avoiding abundance or ionization effects. The higher the upper level of the corresponding transition is, the lower the radius of the line formation region is located in the disk, owing to decreasing oscillator strength. Hence, from Table \ref{tab:detectedlines} we can infer that the azimuthal velocity of both disks decreases with radius, as expected for Keplerian disks.

Identification of forbidden metallic lines is crucial to validate the B[e] classification. Among these, the strongest is \ion{[Fe}{ii]} $1.678 \mu m$, which is blended with \ion{[Fe}{ii]} $1.677 \mu m$ and \ion{Fe}{ii]} $1.679 \mu m$. However, these neighboring features are weak enough not to distort significantly the $1.678 \mu m$ spikes, given that their wavelengths match perfectly the expected radial velocities for both spectra. Another forbidden iron line is present in the Mc70-14 spectrum at $1.745 \mu m$, although its red side cannot be distinguished due to severe telluric contamination and blending with the \ion{Mg}{ii} $1.746 \mu m$ line. The latter features are difficult to spot in the Mc20-16 spectrum, as metallic lines are less intense in general and the SNR ratio is lower at the red end of the H-band range.

Moreover, two emission features remain unidentified at $1.69 \mu m$ and $2.19 \mu m$. Although the former resembles a \ion{Si}{ii} feature that appears in qF 362 and the Pistol Star, we dismiss such identification owing to the wavelength being 6 \AA ~lower and the absence of other strong \ion{Si}{ii} lines that would have to be present.

      \begin{figure*}
	\centering
	\includegraphics[width=8.5cm]{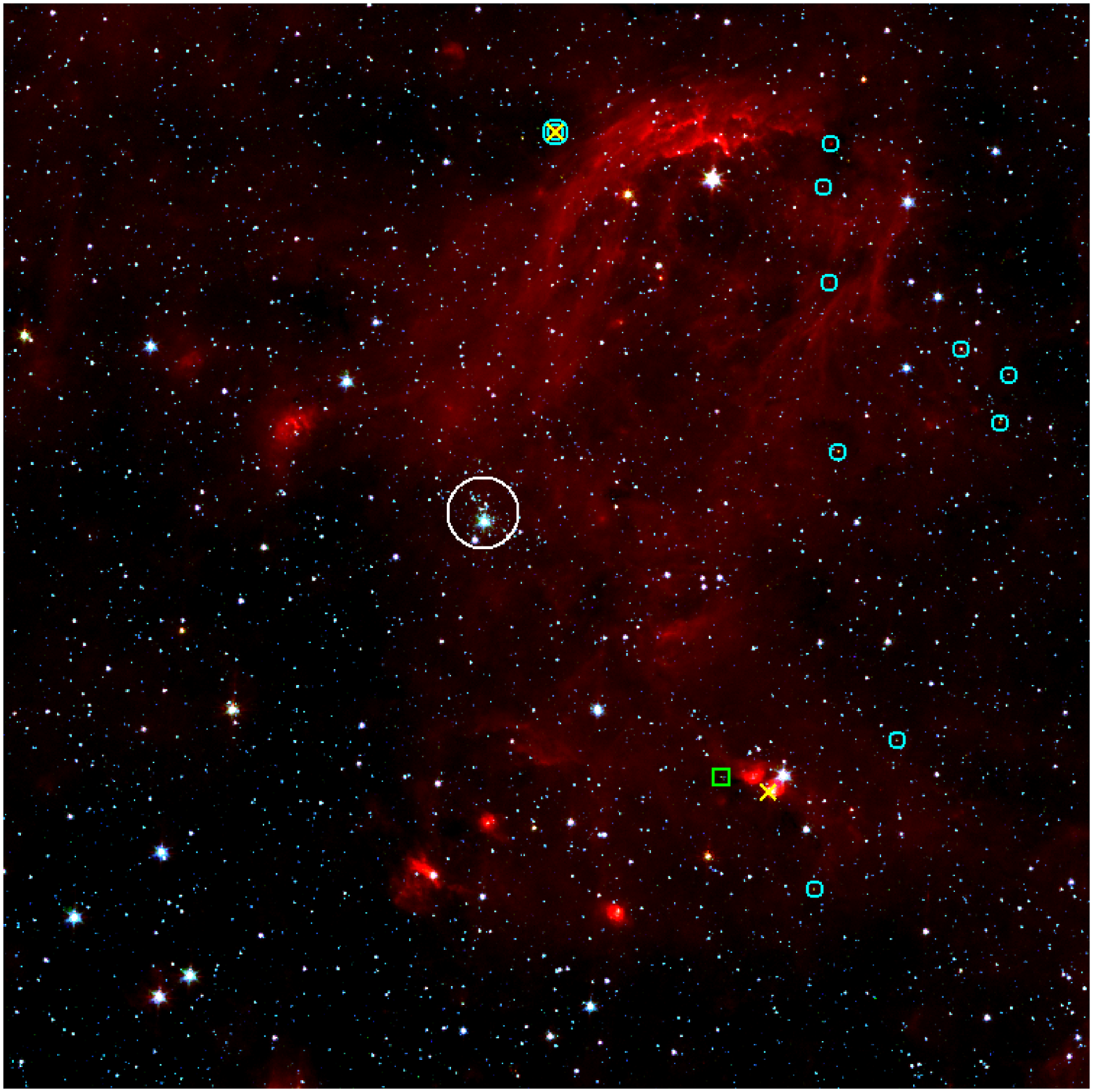}
	~
	\includegraphics[width=8.5cm]{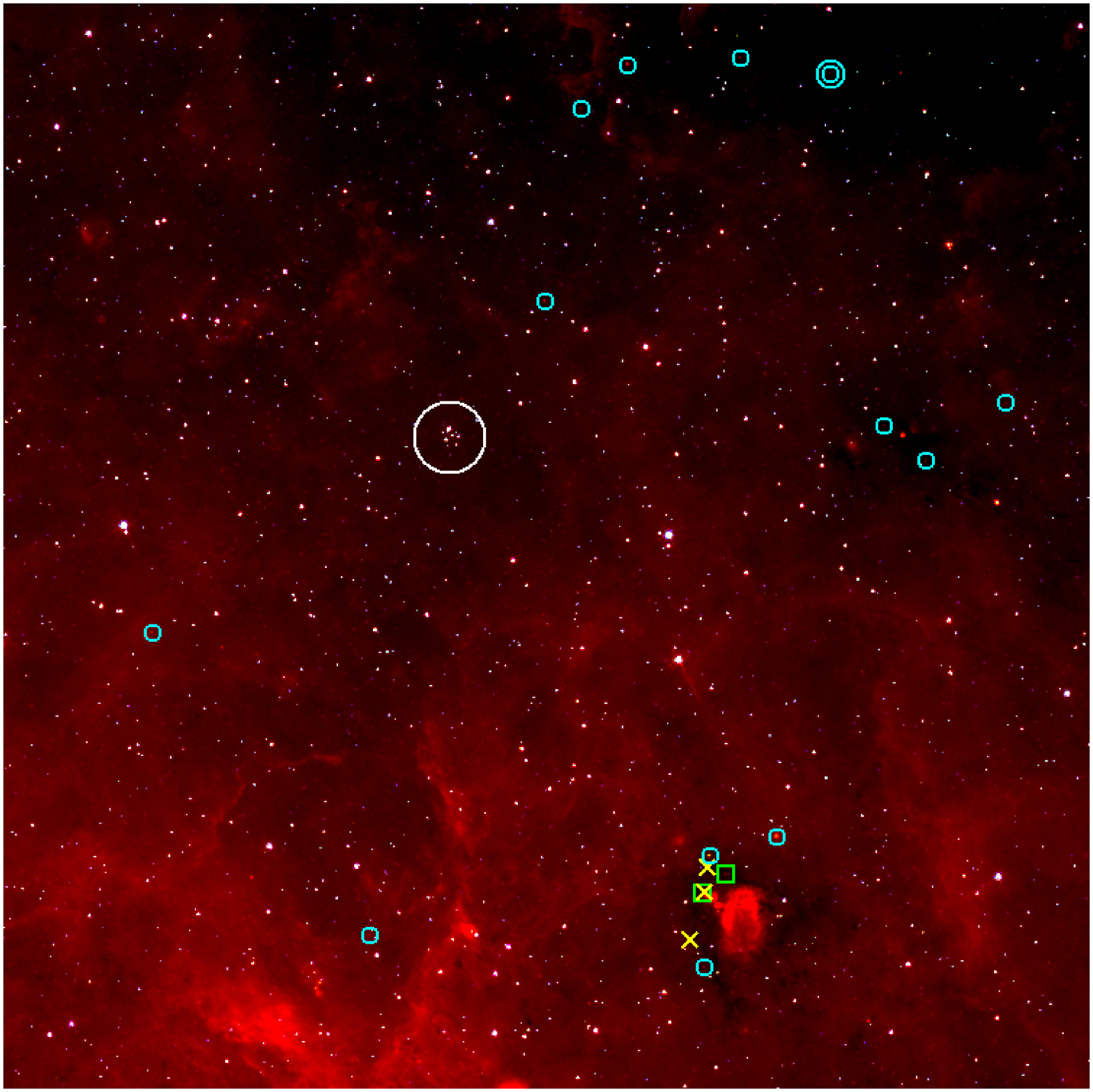}
      
	\caption{$30' \times 30'$ RGB images (with $R=[8.0]$, $G=[4.5]$, $B=[3.6]$) from the GLIMPSE survey showing regions around Mercer 20 (\textit{left}) and Mercer 70 (\textit{right}), which are inside the white circles of radius $1'$. Blue circles are young stellar objects or candidates (double circles: with available distance measurement); green squares poinpoint Extended Green Objects; and yellow crosses are masers. In order to preserve the observational homogeneity in this figure, new YSO candidate detections by \citet{dirienzo+12} have been excluded from the \textit{left pannel} (see text for details). North is up and east is left.}
	\label{fig:GLIMPSEimages}
      \end{figure*}

  \subsection{Neighboring star formation and coevality}

In this section, we aim at discarding membership of Mc20-16 and Mc70-14 to young stellar populations other than the above studied clusters, as well as evaluating the posibility of continuous star formation that invalidates the cluster coevality hypothesis. In any case, the origin of these FS CMa stars must have ages around or below 15$-$18 Myr, as this is roughly the H-burning lifetime of a star whose initial mass is $12 M_{\odot}$ \citep[see e.g.][]{bressan+93,meynet-maeder00,ekstrom+12}.
Hence, we adress these questions by searching for signs of recent star formation in a wide sky area around Mc20 and Mc70. On the one hand, visual inspection of images from the GLIMPSE survey has allowed us to examine the morphology of surrounding star forming regions, mainly revealed by hot molecular emission at the $8 \mu m$ IRAC band. On the other hand, we have looked up objects associated with recent star formation in the SIMBAD Astronomical Database. Fig. \ref{fig:GLIMPSEimages} shows the resulting Young Stellar Objects (YSOs) or candidates, Extended Green Objects (EGOs) and masers, superimposed on quarter square degree GLIMPSE images around the clusters.

Most of the hot molecular emission in the Mc20 surroundings is gathered in the first (west to north) quadrant, on the G044.28+0.11 region. This region consists of a well-defined cometary structure whose tail is oriented toward the cluster and a fainter rim that crosses such tail perpendicularly to its symmetry axis. \citet{dirienzo+12} performed an extensive survey of YSO candidates, where the southern edge of the search region crosses the cluster.
As this survey only covers a partial area of the Mc20 surroundings, we omit the correspoding detections in the left pannel of Fig. \ref{fig:GLIMPSEimages},in order to avoid selection effects between the undersampled and oversampled areas; only sources from \citet{robitaille+08} are kept. However, the class I YSO distribution found by \citet[][see their Fig. 11]{dirienzo+12} is majorly the same than seen in Fig. \ref{fig:GLIMPSEimages} (although in higher numbers): these are concentrated near the cometary region edge. A few more YSO candidates are on the aforementioned fainter rim, which is closer to Mc20; however, these are clearly situated on the rim and are not expected to be associated with the cluster.

On the nortwestern edge of the cometary region near Mc20, \citet{urquhart+09} found an ultra compact (UC) \ion{H}{ii} region (VLA G044.3103+00.0410), associated to a methanol maser \citep[MSX6C G044.3103+00.0416;][]{pandian-goldsmith07}. This UC \ion{H}{ii} region has a measured distance is 8.0 kpc \citep[][and references therein]{urquhart+13}, which is coincident with the Mc20 distance estimated above. On the other hand, \citet{dirienzo+12} found near and far kinematic distances of 4.33 and 7.7 kpc for the cometary region; the latter is strongly favoured by \citet{anderson-bania09} through non-detection of background \ion{H}{I} with the same LSR velocity. Nonetheless, \citet{dirienzo+12} used the ``near'' option to estimate a dynamical cloud age of $t_{\ion{H}{ii}}=2.4 Myr$, by means of the \citet{dyson-williams80} formula:

\begin{equation}
 t_{\ion{H}{ii}}=7.2 \times 10^4\left(\frac{R_{\ion{H}{ii}}}{pc}\right)^{4/3}\left(\frac{Q_{Ly}}{10^{49} s^{-1}}\right)^{-1/4}\left(\frac{n_i}{10^3 cm^{-3}}\right) yr \,,
\end{equation}

where $R_{\ion{H}{ii}}$ is the radius, $Q_{Ly}$ the ionizing luminosity and $n_i$ the gas density. Since calculated radius and luminosity vary with distance as $\sim D$ and $\sim D^2$ respectively, then $t_{\ion{H}{ii}}\propto D^{5/6}$. Therefore, the corrected dynamical age at the far distance would be 3.9 Myr, which is  somewhat lower than the Mc20 age. This is compatible with the triggered creation of the bubble by Mc20 at the time this cluster emerged from the natal cloud. Despite having omitted the presence of Mc20, \citet{dirienzo+12} also concluded a triggered star formation mode for this region, based on physical features and the cometary morphology. This scenario implies that star formation has moved to the region edges, being inhibited at the starting location (i.e. the cluster). The velocity difference between Mc20 and G044.28+0.11 is consistent with the elongated shape of the molecular region and the relocation of the cluster far from the region center.

Additional $8 \mu m$ emission comes from smaller regions that are situated south/southwest of Mc20. The brightest cloud is the [MKJ2009]58 molecular clump \citep{matthews+09}. These authors calculated kinematic distances of 5.1 and 7.1 kpc for the near and far intersections with the galactic rotation curve; the latter would be consistent with being in the same spiral arm than the cluster and the aforementioned cometary structure. Very close to this cloud, two nearby observational signs of ongoing star formation are present: a SiO maser \citep[object ``Onsala 100'' from][]{harju+98} and a EGO \citep[G044.01-0.03; ][]{cyganowski+08}.
\citet{chen+10} gave a kinematic distance of 5.3 kpc for the EGO assuming the ``near'' option; this assumption is probably erroneous and this object is spatially coincident with the molecular clouds in the Mc20 wide field. In any way, none of the small clouds we refer in this paragraph are within a $7'$ radius, therefore we can discard them as either the origin of Mc20-16 or a sign of continuous star formation in the Mc20 field.

In contrast to the Mc20 environment, visual inspection of the M70 surroundings does not reveal any hot dust structure that is clearly related to the cluster. The $8 \mu m$ extended emission is enhanced toward the galactic plane, which is situated 35 arcminutes south of Mc20. The only well-defined clump in the field is the IRAS 15557-5215 \ion{H}{ii} region, hosting two EGOs \citep{cyganowski+08}, three candidate YSOs and three methanol masers.
The latter have the following identifiers and available velocity and/or distance measurements: 329.457+0.506, $\varv_{LSR}=-67.6 ~km~s^{-1}$ \citep{bronfman+96}, $d_{Near}=4.5 kpc$, $d_{Far}=10.1 kpc$ \citep{ellingsen05}; G329.48+0.51, $\varv_{LSR}= -72 ~km~s^{-1}$ \citep{schutte+93}; G329.469+0.502, $\varv_{LSR}=-69.5 ~km~s^{-1}$, $d_{Near}=4.5 kpc$, o $d_{Far}=10.1 kpc$ \citep{valtts+00}. Despite being nearly coincident between them, these are too far away from the Mc20 measurents, therefore such objects correspond to foreground or background clustered star formation, probably in a different Galactic spiral arm.

Several candidate YSOs are scattered over the remaining sky area in the right pannel of Fig. \ref{fig:GLIMPSEimages}, where no clear associations with clouds are visually perceived. One of these objects is the MSXDC G329.67+0.85a dense core, whose velocity is $\varv_{LSR}=-45.8 ~km~s^{-1}$. Despite not having available velocity or distance measurements for the remaining YSO candidates, physical connection with the cluster is very unlikely owing to the angular separations, with the closest one being placed at 4.5 arcminutes from the Mc20 center. Together with the absence of $8 \mu m$ emission on the outskirts of the cluster, all these data allow us to conclude that star formation have been probably stopped around Mc20 and the surroundings have been cleared of gas, while signs of ongoing or formation in the field are foreground or background.

Finally, we cannot totally rule out that Mc20-16 and Mc70-14 have formed in isolation. However, this scenario is highly improbable, given that a vast majority of stars are born in clusters \citep{lada-lada03,portegieszwart+10}. Furthermore, the fraction of OB stars that are located in clusters or running away from them is particularly high, exceeding 95\% \citep{clarke+00,dewit+05,pflammaltenburg-kroupa10,gvaramadze+12}. Such lower limit, together with the absence of additional clustered star formation within several arcminutes, yields a near 100\% probability for Mc20 and Mc70 as the birthplaces of the FS CMa stars presented here.
 
      \begin{figure}
	\centering
	\includegraphics[width=\hsize]{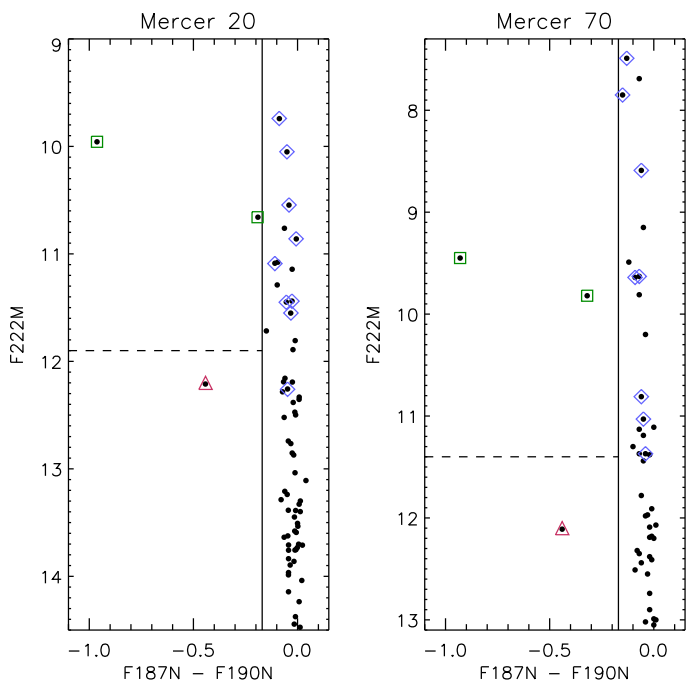}
	\caption{$P_\alpha$-magnitude diagram of the Mc20 and Mc70 fields. Red triangles are FS CMa stars, green squares are WR or ``slash'' stars and blue diamonds are OB cluster members. For the sake of clarity, GLIMPSE20-1 is excluded for being situated far away from the magnitude range drawn here. The solid line shows our adopted limit for  $P_\alpha$ excess and the dashed line is at $M_K \approx -4$.
              }
        \label{fig:cmds}
      \end{figure}
 
  \subsection{Photometric properties}
  \label{sec:FSCMaPhot}
     
As suggested by Figs. \ref{fig:Mc20emissions} and \ref{fig:Mc70emissions}, narrow-band photometry is a key method for distinguishing FS CMa stars among other coeval hot massive stars, especially when a significant fraction of them are much brighter. Among the most luminous hot massive stars, the most easily detected sources through $P_\alpha$ photometry are LBVs and WRs, as shown in the Galactic Center region \citep{wang+10, mauerhan-morris+10, mauerhan-cotera+10}, while extreme OB supergiants only emit weakly in $P_\alpha$. Fig. \ref{fig:cmds} shows a diagram separating the strong $P_\alpha$ emitters from the remaining objects for each cluster; we establish the empirical limit between strong and weak $P_\alpha$ emission at F187N $-$ F190N $\approx -0.17$. The only stars appearing in the diagram region for strong emitters are WR, ``slash'' stars, and FS CMa objects.
Additionally, the latter are also separated by a broad magnitude gap, as expected for their relatively low brightness. The most luminous main sequence star displaying the B[e] phenomenon should have an O9 subtype and $M_K \approx -3.3$ \citep{martins-plez06}. If we duplicate the luminosity of such star to take into account (roughly) the disk contribution, we reach $M_K \sim -4$ This coincides approximately with the K-band magnitude of the least luminous WR stars in the Milky Way \citep{crowther+06}. Using the extinction and distance values above calculated, this limit is equivalent to F222M $\approx 11.9$ and F222M $\approx 11.4$ for Mc20 and M70 respectively. Thus, we have found a differentiated region where FS CMa stars are expected to be present in $P_\alpha$-F222M diagrams of young clusters hosting WR stars.

Since a key feature of FS CMa star is a strong infrared excess with a steep decrease in the mid-infrared, we have explored the available photometric data at longer wavelengths. Unfortunately, stellar crowding in the regions where Mc20-16 and Mc70-14 are located complicates the detection for mid-infrared instruments, whose resolving power is generally lower than in the near-infrared. Specifically, Mc20-16 only appears in the GLIMPSE survey, whose spatial resolution ($\sim 0.2''$) is the highest among the mid-infrared public surveys. On the other hand, Mc70-14 cannot be resolved due to the presence of four higher luminosity stars within a $3''$ radius (see Fig. \ref{fig:Mc70finders}), including the most luminous cluster member (Mc70-1).

      \begin{figure}
	\centering
	\includegraphics[width=\hsize, bb=10 5 275 255, clip]{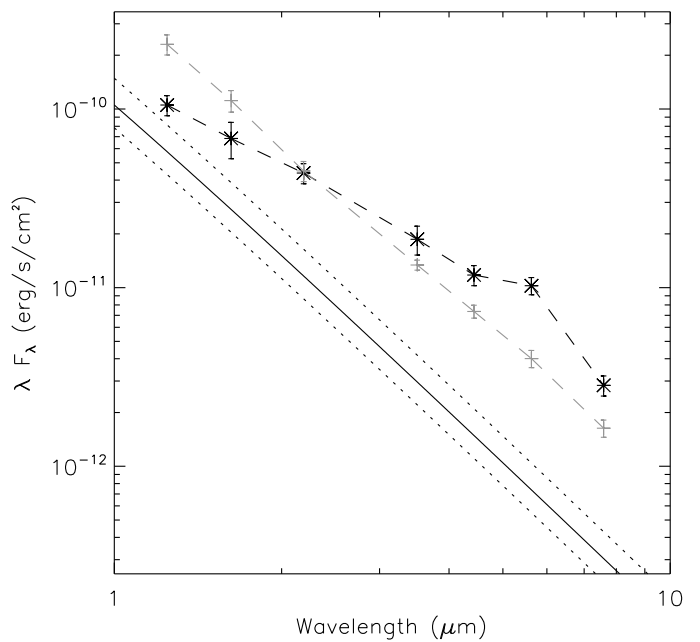}
	\caption{Dereddened SEDs of Mc20-16 (black) and Mc20-17 (grey). A black body of $T = 28,000 K$ and $R = 5 R_\odot$ at a distance of $8.2 \pm 1.3~kpc$ is also plotted (black solid curve; uncertainty is represented by dotted lines).
              }
        \label{fig:sed}
      \end{figure}

In order to build the Spectral Energy Distribution (SED) of Mc20-16 we have derredened the UKIDSS data using $A_\lambda = \lambda^{-(1.94 \pm 0.21)}$ and $A_K = \bar A_K^{Mc20} = 1.15 \pm 0.14$, while we have corrected the extinction in the GLIMPSE data through the \citet{indebetouw+05} law. For comparison purposes, we have also calculated the SED of Mc20-17, whose spectral type (O9-B2 III-V) is nearly the same than the Mc20-16 stellar component, although slightly more luminous. Derredening of Mc20-17 has been carried out similarly to Mc20-16, but using the $A_K^{Mc20-17} = 1.35$ value individually estimated instead of the cluster average. This choice has been preferred as this object being located in the Mc20 outskirts appears significantly redder than other early-B cluster members, which points at a differential extinction effect.

The infrared SEDs of Mc20-16 and Mc20-17 are shown in Fig. \ref{fig:sed}, along with the Planck distribution for a black body of  $T = 28,000 K$ and $R = 5 R_\odot$ at the same distance than the cluster. These values correspond to 12$-$14 $M_\odot$ models of 5$-$6 Myr old \citep{ekstrom+12}, with small variations depending on the rotational velocity and the exact age; a normal B0.5-1 V star would fulfill such conditions \citep{pecaut-mamajek13}. As expected, the Mc20-17 and black body slopes are nearly coincident, while the Mc20-16 SED shows a remarkable infrared excess that peaks at the [5.8] IRAC band. The clearly seen decrease at the [8.0] band implies a very compact and warm dusty disk. Towards shorter wavelengths, the Mc20-16 flux density approaches to the Planck distribution depicted in Fig.\ref{fig:sed}, allowing us to confirm that the radius and luminosity of the underlying star are what we would expect from an early-B main-sequence star.

Due to the unavailability of reliable photometric data of Mc70-14 from public surveys, we can only compare it with the Mc20-16 photometry through the NICMOS data. The observed F160W $-$ F222M color excesses for these objects are $E^{Mc20-16} = 1.37 \pm 0.05$ and $E^{Mc70-14} = 1.06 \pm 0.05$, which include both the interstellar extinction and the disk effects. The location of these stars in the central cluster regions allows us to separate such contributions to the color excess under the assumption of undergoing the same extinction than the average for each cluster, within errors. These averages are $E_{interst}^{Mc20} = 0.93 \pm 0.12$ and $E_{interst}^{Mc70} = 0.79 \pm 0.16$. Therefore, the color excesses in the F160W $-$ F222M band caused by the disks are $E_{disk}^{Mc20-16} = 0.44 \pm 0.17$ and $E_{disk}^{Mc70-14} = 0.27 \pm 0.21$. Due to the relatively high uncertainties, we can only conclude that Mc70-14 has a positive near-infrared excess that is equal or somewhat lower than the  Mc20-16 excess.

Mc20-16 and Mc70-14 have the weakest infrared excesses among the few FS CMa stars with measured SEDs, as can be seen by comparing Fig. \ref{fig:sed} with \citet[][Fig.20]{miroshnichenko+07} and \citet{miroshnichenko+11a}. Such weakness is not entirely unexpected since the stellar components are among the hottest objects in the FS CMa class, and therefore dust is more efficiently destroyed by the more energetic radiation of the central body. However, the presence of multiple hot luminous stars at sub-parsec distances is an additional contribution of great importance here. As shown by \citet{hollenbach-adams04}, photoevaporation caused by UV radiation from nearby massive stars has a strong effect on dispersal of disks in young clusters, and this mechanism becomes predominant in the outer regions of the disks. The latter also could explain the aforementioned SED decrease even at shorter wavelengths (6$-$8 $\mu m$) than for other FS CMa stars \citep[10$-$30 $\mu m$;][]{miroshnichenko07,miroshnichenko+07}.

Cluster membership of FS CMa stars also allows us to calculate absolute magnitudes for this kind of objects in an unprecedented way. We use the extinction and distance estimates from previous subsections, however we cannot utilize the NICMOS/UKIDSS photometric transformation that was only valid for ``normal'' OB stars (see section \ref{sec:photometry}). Given that both FS CMa stars have very similar spectra and F160W $-$ F222M colors, we prefer to take $(F222M-K) = 0.30$ from Mc20-16 and apply it to Mc70-14. Thus, the resulting K-band absolute magnitudes are $M^{Mc20-16}_K = -3.8 \pm 0.5$ and $M^{Mc70-14}_K = -3.4 \pm 0.4$.
These values correspond to the whole FS CMa objects, i.e. the total outgoing flux that include not only the star + disk luminosity, but also the effect of the circumstellar extinction. On the one hand, this implies the K-band luminosity of the stellar component must be lower; the Mc20-16 SED (Fig. \ref{fig:sed}) provides order-of-magnitude estimate for the K-band excess of $\sim 1~mag$, therefore  $M^{stellar}_K \sim -2.5$. On the other hand, the $0.4~mag$ diference between both stars can be compensated if we consider the posibility of a greater excess for Mc20-16, as suggested previously in this section. Hence, we state that both stellar components virtually have the same luminosity within the associated uncertainties. 

\begin{table*}
  \caption{Equatorial coordinates and NICMOS/HST photometry of new FS CMa candidates.}             
  \label{tab:candidates}      
  \centering          
  \begin{tabular}{c c c c c c c c}       
  \hline\hline       			
 Cluster   &  Star 	& 	R.A. 		& 	Dec 		  & $m_{F160W}$	& $m_{F222M}$	& $P_\alpha$	& $m_{F222M}$(WR/slash)\tablefootmark{a}\\ 
  \hline
 Mercer 81 &  Mc81-28   &  $16^h 40^m 30.73^s$  &  $-46\degr 23' 17.8''$  &	16.05   &	13.88	&	-0.26	&	[10.59,11.46]	\\
 Danks1	   &  D1-13     &  $13^h 12^m 29.23^s$  &  $-62\degr 42' 06.9''$  &	11.89	&	11.28	&	-0.20	&	[6.62,8.31]	\\ 
 Danks1    &  D1-14     &  $13^h 12^m 26.98^s$  &  $-62\degr 42' 03.9''$  &	11.93	& 	11.37	&	-0.27	&	[6.62,8.31]	\\ 
\hline                  
\end{tabular}
\tablefoot{\tablefoottext{a}{F222M magnitude ranges of Wolf-Rayet and ``slash'' members in each corresponding cluster.}}
\end{table*}

  \subsection{A search for new FS CMa candidates}

Extensive surveys for finding new FS CMa stars in clusters are beyond the scope of this paper. More modestly, we have simply explored the available Paschen-$\alpha$ photometry of young clusters from our own NICMOS/HST data in the \#11545 program. Specifically, we have only examined clusters whose evolutionary stages are similar to Mc20 and Mc70 (e.g. with WR of blue supergiant cluster members) in order to apply the $P_\alpha$ photometric criterion (Fig. \ref{fig:cmds}) to FS CMa candidate selection. Such clusters are Mercer 23 \citep[previously studied by][]{hanson+10}, Mercer 30 \citep{kurtev+07}, Mercer 81 \citep{mc81_1,mc81_2}, Danks1 and Danks2 \citep{davies+12}.

This search yields three new candidates whose coordinates and NICMOS/HST photometric measurements are listed in Table \ref{tab:candidates}. F222M magnitude ranges of WR and ``slash'' stars in the corresponding clusters are provided for comparison. As expected, these evolved stars are between 2 and 5 magnitudes brighter than the FS CMa candidates, which is expected given that Mc20-16 and Mc70-14 are near the high-luminosity limit of FS CMa stars and a magnitude range of $\Delta m_{F222M} \approx 5$ is allowed for such candidates.  
      
      \begin{figure}
	\centering
	\includegraphics[width=5cm, bb=40 15 435 395, clip]{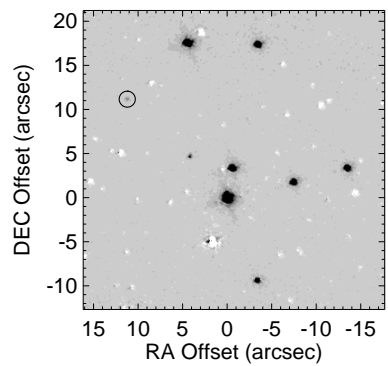}
	\caption{F187N $-$ F190N subtraction image of Mercer 81, showing only the region where emission-line stars are placed. The complete image was published by \citet{mc81_1}, pinpointing all the $P_\alpha$ emitters except Mc81-28 (circled here), which was unperceived due to its relative faintness.}
        \label{fig:mc81}
      \end{figure}

Despite its strong Paschen-$\alpha$ emission, Mc81-28 was previosly overlooked, partly due to the relative faintness when compared with bright cluster members \ref{fig:mc81}. Moreover, this object does not appear in the corresponding $P_\alpha$-F222M diagram \citep[see][Fig.4, central pannel]{mc81_1} given that this star is situated marginally outside the covered sky area (15 arcsec from the cluster center). On the other hand, D1-13 and D1-14 are clearly shown by \citet[][Fig. 5, right pannel]{davies+12}.

\section{Discussion}

We have found the following evidence suggesting that the newly-discovered FS CMa stars are part of two young coeval clustered populations. First, spectral types of the Mc20-16 and Mc70-14 underlying stars are among expected in populations of a few Myr old, being highly improbable that unrelated stars of the same type are projected close to the cluster centers. Second, radial velocity results that were calculated in section \ref{sec:2fscma} are consistent with the cluster values. Third, neither the clusters nor the neighboring regions present signs of continuous star formation that allows to discard coevality. Finally, we have dismissed the existence of unrelated regions of recent star formation that are simultaneously coincident with the radial velocities and lines of view toward Mc20 and Mc70. Despite each individual evidence leaves room for alternative possibilities, we consider the complete set as a concluding proof of membership to coeval clusters.

Cluster membership and coevality allow us to assign an age range from 3.5 to 6.5 Myr to the newly-discovered FS CMa stars. The lower limit excludes the possibility of a pre-main sequence origin of the disks, given that such phase lasts $\sim 0.1 Myr$ for $12 M_\odot$ objects \citep{bressan+12}. The upper limit allows us to rule out post-main sequence evolutionary stages where the B[e] phenomenon can be shown (i.e. B[e] supergiants and protoplanetary nebulae), as the Geneva models \citep{ekstrom+12} yield a 15-20 Myr main-sequence lifetime for $12 M_\odot$ stars. Although the age range does not forbid completely the existence of a symbiotic B[e] star, the cool giant/supergiant companion would be much brighter in the near-infrared and its spectral features would be strongly visible in H and K bands.
Therefore, the B[e] stars we have detected in Mc20 and Mc70 must belong to the FS CMa subclass. Moreover, all the observational and physical criteria that have been measured or estimated in this paper for Mc20-16 and Mc70-14 fulfill the definition of FS CMa stars \citep{miroshnichenko07}, which include the existence of the B[e] phenomenon, an spectral type between O9 and A2, and a relatively low luminosity ($2.5 \lesssim log(L/L_\odot) \lesssim 4.5$). Also, the infrared excess is consistent with a compact and warm dusty disk, as expected for FS CMa stars.

At this point, it is necessary to remark that evolutionary models that are cited in several sections of this paper correspond to single star evolution. Hence, stating that ages, luminosities and temperatures of Mc20-16 and Mc70-14 are compatible with main-sequence stars does not imply that these are genuine main-sequence objects, if we admit a binary evolution scenario. In particular, the 12-14 $M_\odot$ model to which we referred in section \ref{sec:FSCMaPhot} must be only taken for comparison purposes; the actual mass of the Mc20-16 stellar component may be somewhat different (e.g. for a merger case, see below). Nonetheless, moderate mass changes do not affect the reasoning about the evolutionary state of Mc20-16 and Mc70-14, since pre-main sequence objects are expected only for $M \lesssim 2M_\odot$ \citep{bressan+12} and evolved stars for $M \gtrsim 32M_\odot$ \citep{ekstrom+12} at the cluster ages.

The above presented observations provide new clues for the binarity-related hypotheses that were summarized in the introduction of this paper. Particularly, the posibility of FS CMa stars being intermediate-mass post-AGB binaries \citep{miroshnichenko+13} is completely elliminated by the confirmed ages of Mc20-16 and Mc70-14. Despite our spectra (Fig. \ref{fig:fscmaspec}) do not show any companion sign, we cannot dismiss other hypothesis involving binarity. Cooler main-sequence companions may exist, provided that the luminosity ratios are high enough to have the spectral features of the secondary star below the noise level.
Since SNR $\approx$ 100-150, this entails a lower limit of $\sim 15$ for the luminosity ratio in the near-infrared, or equivalently, $\Delta K \gtrsim 3$. In principle, such condition would restrict the companion to spectral types later than B7 V, however this constraint moves to earlier subtypes ($\sim$ B5) owing to the fact that the only strong features in mid-B main-sequence stars (H, \ion{He}{I}) are coincident with the strongest lines in B[e] stars, being harder to separate both contributions.

Interestingly, the locations of both Mc20-16 and Mc70-14 just in the most crowded regions of the host clusters supports the binary origin hypothesis. Close encounters between a tight binary and other cluster members make the binary orbit tend to be tighter and more eccentric \citep{heggie75, portegieszwart+10}, while stellar crowding enhances the encounter rate. On the one hand, this favours the recent mass transfer scenario that is caused by a close secondary star in an eccentric orbit \citep{millour+09}. On the other hand, these frequent dynamical interactions can eventually facilitate the occurrence of a merger.
Mergers have been proposed to explain the formation of B[e] supergiants \citep{pasquali+00,podsiadlowski+06}, likewise a lower luminosity version of this mechanism could produce FS CMa stars. In order to evaluate this new hypothesis, it is mandatory to refer to V1309 Sco, which is the only fully documented case of a stellar merger \citep{tylenda+11}. Post-merger infrared observations by \citet{nicholls+13}, along with simulations by \citet{nandez+14}, and dust modeling by \citet{zhu+13}, proved that a great amount of matter was expelled during the merging event and was converted into dust grains soon after. Besides this, simulations of mergers whose primary component is a main-sequence massive star \citep{glebbeek+13} yield a stellar product with a helium-enhanced core.
As a consequence, the final star seems to be another main sequence star that behaves as if it were slightly more massive than it actually is, having an increased radius, higher luminosity and temperature and a lower main-sequence lifetime. Hence, FS CMa stars could be post-merger products that still retain part of the dusty ejection in the shape of a disk. In principle, the upper limit for the rotational velocity of Mc70-14 (see section \ref{sec:2fscma}) advises against a coalescence scenario, given that the merger products are expected to rotate very rapidly \citep{demink+13}. However, the combination of the loss of angular momentum \citep[up to 30~\%;][]{nandez+14} caused by mass ejection during the merging event and the yet to be quantified magnetic braking \citep{demink+14} may spin down the star to its observed rotation value.

In order to assess the role of stellar clustering in the origin of FS CMa stars (e.g. binary evolution), we must wonder how often these objects are located in clusters like Mc20 and Mc70. In principle, the detection of two FS CMa stars plus three candidates in a sample of seven clusters of similar ages (a few Myr) and masses ($\sim 10^4 M_\odot$) favours a relatively high frequency in this kind of young massive clusters. Then, why FS CMa stars have not been found in clusters until now? This apparent paradox could be caused by an observational bias. If clusters are required to be roughly as massive as Mc20 or Mc70 to have a significant probability to host one FS CMa star, then the latter will be totally outranked (in terms of luminosity) by many cluster members that are more massive or more evolved, depending on the age. Additionally, as most of young massive clusters are distant and highly reddened, spectroscopy is usually limited for the brightest targets.

This condition is already evident for Mc20-16 and Mc20-70 and the host clusters, although these two objects are close to the upper luminosity limit expected for FS CMa stars. In fact, our research was initially aimed at observing the most massive evolved cluster members, which were pinpointed by their typical $P_\alpha$ emission (e.g. WR, LBV), and only the unexpected finding of less luminous emission-line stars led us to select them for spectroscopic observations, in spite of being time-consuming for the telescope.

In any way, two confirmed cases are not sufficient to infer the frequency of FS CMa stars in clusters or understand their dependance on the environment. A systematic survey for new FS CMa stars in young clusters is crucial to answer all these questions, and the photometric criterion we have suggested in section \ref{sec:FSCMaPhot} could be very useful for such search. 

\section{Summary}

In this paper, we have reported the first detections of FS CMa stars in clusters. The presence of such B[e] objects in coeval populations of massive stars has also allowed us to constrain some of their observational properties in an unprecedented way. Our scientific strategy and outcomes are summarized in the following points:

   \begin{enumerate}
    
    \item Based on new spectroscopic data of massive stars in Mercer 20 and Mercer 70, we have considerably improved the knowledge of these massive ($\sim 10^4 M_\odot$) clusters in several aspects. We have identified and classified 22 new cluster members, including a very compact group of three objects that were previously misidentified as a single Wolf-Rayet star. More importantly, extinction, radial velocity and distance have been accurately calculated for both clusters, while ages have been estimated by means of state-of-the-art evolutionary models.
    
    \item Spectra of Mc20-16 and Mc70-14 have lead to an early-B[e] classification and have allowed to confirm the presence of rotating circumstellar disks.
    
    \item Membership of these B[e] objects to their respective clusters, together with coevality, have been investigated thoroughly. Putting together our cluster characterization and literature data of neighboring regions, we have found neither signs of continuous star formation nor alternative origins of the newly discovered objects. Conversely, spectral types and radial velocities of Mc20-16 and Mc70-14, along with their locations in the central regions of the clusters, form enough evidence to assure that these objects are part of the coeval populations of Mc20 and Mc70.
    
    \item Since cluster membership and coevality legitimizes the direct application of cluster results to the corresponding constituents, we have made use of interstellar extinction, distances and ages to constrain the nature and evolutionary state of Mc20-16 and Mc70-14. On the one hand, ages ($\sim 6$ and $\sim 4.5$ Myr, respectively) have allowed to confirm the FS CMa classification, discarding other B[e] subclasses. On the other hand, we have utilized cluster distances and interstellar extinction results to constrain the photometric properties of these stars and their circumstellar disks, which coincide with the expected attributes of FS CMa stars. Both ages and luminosities are congruent with a main-sequence evolutionary state.
    
    \item As we have demonstrated, the characteristic Paschen-$\alpha$ emission of FS CMa stars is an ideal tool for pinpointing them easily, even in crowded fields where these objects are relatively faint. Consequently, we have proposed a new method to find FS CMa candidates in young clusters based on narrow-band photometry. After testing this method on available photometric data, we have found three new candidates in the young massive clusters Mercer 81 and Danks 1.
    
    \item We have discussed the aftermath of FS CMa discoveries in young clusters, especially regarding their origin and evolutionary state. We have dismissed the post-AGB binary nature of FS CMa stars and have explored other binarity-related possibilities, including a new hypothesis about post-merger objects.
    
    \end{enumerate}
    
Finally, we remark that new detections of FS CMa in clusters are needed to adress unsolved questions about the influence of clustered environments in the creation of the B[e] phenomenon in main-sequence-like objects.

\begin{acknowledgements}

We wish to thank the anonymous referee for the useful suggestions to this paper. We are also grateful to Miriam Garcia for estimating the rotational velocities and stimulating discussions. We would like to thank Ignacio Negueruela and Alejandro B\'aez-Rubio for helpful discussions about FS CMa stars. Part of this research has been supported by the Spanish Government through projects AYA2010-21697-C05-01, FIS2012-39162-C06-01 and ESP2013-47809-C3-1-R. D. dF. also acknowledges the FPI-MICINN predoctoral grant BES-2009-027786. This paper is partly based on observations made with the NASA/ESA Hubble Space Telescope, obtained at the Space Telescope Science Institute, which is operated by the Association of Universities for Research in Astronomy, Inc., under NASA contract NAS 5-26555; these observations are associated with program \#11545. This work is based in part on data obtained as part of the UKIRT Infrared Deep Sky Survey.
This research has made use of the SIMBAD database, operated at CDS, Strasbourg, France. This work is based in part on observations made with the Spitzer Space Telescope, which is operated by the Jet Propulsion Laboratory, California Institute of Technology under a contract with NASA. This publication is also based on data obtained from the ESO Science Archive Facility under request numbers DDELAFUENTE51468, 51471, 51496, 60232.
\end{acknowledgements}


\bibliographystyle{aa} 
\bibliography{paper1} 

\end{document}